\documentclass[aps,onecolumn,preprintnumbers,showpacs,showkeys,nofootinbib,
superscriptaddress  
]{revtex4}

\usepackage{epsfig}
\usepackage{amssymb,amsmath,amsfonts,amsthm,graphicx,psfrag}

\newcommand{\tr}{\operatorname{Tr}}
\newcommand{\beq}{\begin{eqnarray}}
\newcommand{\eeq}{\end{eqnarray}}

\begin{document}
\hfill{\bf ITEP-LAT-2015-01, HU-EP-15/12}

\title{Two-Color QCD with Non-zero Chiral Chemical Potential}
\author{V.~V.~ Braguta}
\affiliation{Institute for High Energy Physics, 142281 Protvino, Russia}
\affiliation{Far Eastern Federal University,  School of Natural Sciences, 
690950 Vladivostok, Russia} 
\author{V.~A.~Goy}
\affiliation{Far Eastern Federal University, School of Biomedicine, 
690950 Vladivostok, Russia}  
\author{E.-M.~Ilgenfritz}
\affiliation{Joint Institute for Nuclear Research, BLTP, 141980 Dubna, Russia}
\author{A.~Yu.~Kotov}
\affiliation{Institute of Theoretical and Experimental Physics, 
  117259 Moscow, Russia}
\author{A.~V.~Molochkov}
\affiliation{Far Eastern Federal University, School of Biomedicine, 
  690950 Vladivostok, Russia}
\author{M.~M\"uller-Preussker}
\affiliation{Humboldt-Universit\"at zu Berlin, Institut f\"ur Physik,
12489 Berlin, Germany}
\author{B.~Petersson}
\affiliation{Humboldt-Universit\"at zu Berlin, Institut f\"ur Physik,
12489 Berlin, Germany}

\begin{abstract}

The phase diagram of two-color QCD with non-zero chiral chemical potential is 
studied 
by means of lattice simulation. 
We focus on the influence of a chiral chemical potential on the 
confinement/deconfinement phase transition and the breaking/restoration 
of chiral symmetry. The simulation is carried out with dynamical staggered 
fermions without rooting. The dependences 
of the Polyakov loop, the chiral 
condensate and the corresponding susceptibilities on the chiral chemical 
potential and the temperature are presented. The critical temperature is 
observed to increase with increasing chiral chemical potential. 
\end{abstract}

\pacs{12.38.Aw, 12.38.Gc, 11.15.Ha}
\keywords{Lattice QCD, staggered fermions, non-zero temperature,
chiral chemical potential, chiral condensate}

\maketitle

\section{Introduction}

For the vacuum state of QCD as well as for the properties of finite-temperature 
QCD the existence of non-trivial topological excitations is important.
Well known are instantons~\cite{Belavin:1975fg} as classical solutions 
in Euclidean space as tunneling events between topologically different 
zero-temperature vacua. The role of topology for the solution of the famous 
$U_A(1)$ problem has been 
recognized very early~\cite{Witten:1979vv,Veneziano:1979ec}. 
Presently, the anomalous breaking
of the $U_A(1)$ symmetry above the deconfinement transition/crossover is under
intense investigation in the lattice community (see, e.g.,
Ref.~\cite{Dick:2015twa}).

It is known by now from lattice QCD at zero temperature that a (fractal) 
low-dimensional (laminar) topological vacuum structure is discernible at 
very fine
resolution scale~\cite{Ilgenfritz:2008ia}, while localized instanton-like 
structures, actually prevailing at an infrared scale, are believed to explain 
chiral symmetry breaking~\cite{Schafer:1996wv,Shuryak:1997qb}.

The gluon fields contributing to the path integral at finite temperature
correspondingly may contain calorons~\cite{Kraan:1998pm,Lee:1998bb}.
Adapted to the non-trivial holonomy they have a richer structure (in terms 
of ``dyonic'' constituents) than instantons. The changes of this structure 
at the QCD phase transition are presently under 
study~\cite{Ilgenfritz:2013oda,Bornyakov:2014esa}.

Some time ago the gluonic topological structure and the famous axial anomaly 
have been proposed to be immediately observable (and controllable) through 
the generation of $P$ and $CP$ violating domains (violating also translational 
invariance) in heavy ion collisions~\cite{Kharzeev:2004ey,Fukushima:2008xe}. 
It has been demonstrated by detailed numerical calculations
\cite{Kharzeev:2004ey,Kharzeev:2001ev} that macroscopic domains of 
(anti)parallel color-electric and color-magnetic field can emerge in a heavy 
ion collision creating an increasing chiral imbalance among the quarks which 
are deconfined due to the high energy density. 
In this situation, the magnetic field created by the spectator nucleons
may initiate a charge separation relative to the reaction plane (parallel
to the electro-magnetic field)~\cite{Kharzeev:2007jp}. 
The resulting charge asymmetry of quarks would become observable in terms of 
recombined hadrons (chiral-magnetic effect)~\cite{Abelev:2009ac,Abelev:2009ad}.
The strength (and particularly the dependence on the collision energy) of 
this effect has been theoretically studied and proposed to be a signal of the 
transient existence of liberated 
quarks~\cite{Kharzeev:2004ey,Fukushima:2008xe,Adamczyk:2014mzf}.

In recent years the dependence of the chiral and deconfinement transitions
on the magnetic field has been investigated both in models and ab-initio
lattice simulations, see e.g. ~\cite{Shovkovy:2012zn,D'Elia:2012tr}. It 
remains an open question whether the phase transition from quarks to hadrons,
i.e. the onset of confinement and chiral symmetry breaking (and vice versa), 
depends on the chiral imbalance.

In this paper we study the change of the phase structure induced by a chiral 
chemical potential in an equilibrium lattice simulation. We mimic the 
topological content (of a $CP$ violating, topologically nontrivial gluonic 
background in a heavy ion collision event) by inducing a chiral imbalance, 
which is provided (i.e. frozen) by a non-zero chiral chemical potential. 
In this setting, the modification of the phase diagram by the chiral chemical 
potential $\mu_5$ has been studied mainly by means of effective 
models~\cite{Fukushima:2010fe,Chernodub:2011fr,Gatto:2011wc,Andrianov:2013dta,Andrianov:2013qta, Chao:2013qpa,Yu:2014sla}, 
the predictions of which will later be compared with our results.

On the lattice, contrary to the case of non-zero baryon chemical potential,
simulations with non-vanishing $\mu_5$ are not hampered by a sign 
problem\footnote{In $SU(2)$ there is no sign problem even in presence of 
a baryon chemical potential.}.
Thus, one can employ the standard hybrid Monte Carlo algorithms. 
Such lattice simulations with $\mu_5 \ne 0$ were already performed in 
Ref.~\cite{Yamamoto:2011gk,Yamamoto:2011ks}. The main goal of these papers, 
however, was the study of the chiral magnetic effect. Therefore, the phase 
diagram was not systematically studied.
    
In Refs.~\cite{Braguta:2015sqa,Braguta:2014ira} we have carried out the first 
lattice study of the phase diagram with non-zero chiral chemical potential. 
It was performed in $SU(2)$ QCD with four flavors, which we have considered 
as a simplified model of QCD. In this article we extend this investigation 
generating considerably more data and performing a more detailed analysis.
In this paper we do not attempt to analyse the topological structure that
would reflect the presence of the chiral chemical potential.
  
One reason for choosing the $SU(2)$ gauge group is that less computational 
resources are required for this pilot study than for full QCD. The second 
reason is that we have already carried out two-colour QCD computations with 
an external magnetic field~\cite{Ilgenfritz:2012fw,Ilgenfritz:2013ara}. 
Furthermore, the four flavor case results from our choice of staggered fermions 
as lattice regularization while we avoid to take the root of the fermion 
determinant, which would allow to reduce the number of flavors. 
The ``rooting procedure'' might introduce further systematic errors at finite 
lattice spacing.

In section II we introduce the model and its lattice implementation and define 
the quantities we measure. In section III we present our results, and section 
IV is devoted to their discussion and to the formulation of conclusions. 
In the Appendix we discuss the question of renormalizations refering to 
explicit analytical calculations in perturbation theory.
 
\section{Details of the simulations}

We have performed simulations with the $SU(2)$ gauge group. We employ 
the standard Wilson plaquette action
\begin{equation}\begin{split}
S_g = \beta\sum\limits_{x,\mu<\nu}&\left(1-\frac{1}{N_c}\tr U_{\mu\nu}(x)\right).
\end{split}\end{equation}
For the fermionic part of the action we use staggered fermions 
\begin{equation}\begin{split}
S_f&=ma\sum_x {\bar \psi_x}\psi_x+\\
&+\frac12\sum_{x\mu} 
   \eta_{\mu}(x)({\bar \psi_{x+\mu}U_{\mu}(x)}\psi_x
   -{\bar \psi_x}U^{\dag}_{\mu}(x) \psi_{x+\mu})+\\
   &+\frac12\mu_5a\sum_x s(x)({\bar \psi}_{x+\delta}{\bar U}_{x+\delta, x}\psi_x
   -{\bar \psi}_{x}{\bar U}_{x+\delta, x}^{\dag}\psi_{x+\delta}),
\label{eq:staggeredaction}
\end{split}\end{equation}
where the $\eta_{\mu}(x)$ are the standard staggered phase factors: 
$\eta_1(x)=1,\eta_{\mu}(x)=(-1)^{x_1+\ldots+x_{\mu-1}}$ for $\mu=2,3,4$. 
The lattice spacing is denoted by $a$, the bare fermion mass by $m$,
and $\mu_5$ is the value of the chiral chemical potential.
In the chirality breaking term $s(x)=(-1)^{x_2}$, $\delta=(1,1,1,0)$ 
represents a shift to the diagonally opposite site in a spatial $2^3$ 
elementary cube. The combination of three links connecting sites $x$ and 
$x+\delta$,
\begin{equation}\begin{split}
{\bar U}_{x+\delta,x}=\frac16\sum\limits_{i,j,k=
\text{perm}(1,2,3)}U_i(x+e_j+e_k)U_j(x+e_k)U_k(x) 
\end{split}\end{equation}
is symmetrized over the $6$ shortest paths between these sites. In the 
partition function, after formally integrating over the fermions, one 
obtains the corresponding determinant. As mentioned above, we do not take 
the fourth root of this determinant in order to represent each flavor 
(``taste'') independently of the others. Thus, the continuum limit of our 
model corresponds to a theory of four (degenerate) flavors.

In the continuum limit Eq. (\ref{eq:staggeredaction}) can be rewritten in the
Dirac spinor-flavor basis \cite{KlubergStern:1983dg,MontvayMuenster:2000} as 
follows  
\begin{equation}\begin{split}
S_f \to S^{(cont)}_f = \int d^4x \sum_{i=1}^4 \bar{q_i}
(\partial_{\mu}\gamma_{\mu} +igA_{\mu}\gamma_{\mu}+m+\mu_5\gamma_5\gamma_4)q_i.
\end{split}\end{equation}

We would like to emphasize that the chiral chemical potential, introduced 
in Eq. (\ref{eq:staggeredaction}), corresponds to the taste-singlet operator 
$\gamma_5\gamma_4\otimes\boldsymbol{1}$ in the continuum limit.

It should be also noted here that the usual baryonic chemical potential 
after the discussion in Ref.~\cite{Hasenfratz:1983ba}, 
and also the chiral chemical potential as used in Ref.~\cite{Yamamoto:2011ks}, 
are introduced to the action as a modification of the temporal links by 
corresponding exponential factors in order to eliminate chemical-potential 
dependent quadratic divergencies. For staggered fermions this modification 
can be performed as well for the baryonic chemical potential. However, 
for the chiral chemical potential such a modification would lead to a highly 
non-local action~\cite{Yamamoto:2011ks}. 
Therefore, we decided to introduce $\mu_5$ in Eq.~(\ref{eq:staggeredaction}) 
in an additive way similar to the mass term.

It is known that the additive way of introducing the chemical potential leads 
to additional divergencies in observables. In the Appendix we present 
analytical and numerical investigations of additional divergencies in the 
Polyakov loop and the chiral condensate. Our study shows that there is no 
additional divergency in the Polyakov loop, whereas there is an additional 
logarithmic divergency in the chiral condensate. The latter is numerically 
small and does not effect the results of this paper.

We have performed simulations with two lattice sizes 
$N_{\tau}\times N_{\sigma}^3=6\times20^3, 10\times28^3$. 
The measured observables are
\begin{itemize}
\item the Polyakov loop 
\begin{equation}\begin{split}
L=\frac{1}{N_\sigma^3}\sum_{n_1,n_2,n_3}
\langle\tr\prod\limits_{n_4=1}^{N_{\tau}}U_4(n_1,n_2,n_3,n_4)\rangle \,,
\end{split}\end{equation}
\item the chiral condensate
\begin{equation}\begin{split}
a^3\langle\bar{\psi}\psi\rangle  = 
-\frac{1}{N_{\tau}N_{\sigma}^3}\frac14\frac{\partial}{\partial(ma)}\log Z = 
\frac{1}{N_{\tau}N_{\sigma}^3}\frac14\langle \tr \frac{1}{D+ma} \rangle \,,
\end{split}\end{equation}
\item the Polyakov loop susceptibility
\begin{equation}\begin{split}
\chi_L=N_{\sigma}^3\left(\langle L^2\rangle-\langle L\rangle^2\right)\,,
\end{split}\end{equation}
\item the disconnected part of the chiral susceptibility
\begin{equation}\begin{split}
\chi_{disc}=\frac{1}{N_{\tau}N_{\sigma}^3}\frac1{16}
 (\langle (\tr \frac{1}{D+ma})^2 \rangle-\langle \tr \frac{1}{D+ma} \rangle^2)\,.
\end{split}\end{equation}
\end{itemize} 
The Polyakov loop and the corresponding susceptibility are sensitive to 
the confinement/deconfinement phase transition, whereas the chiral condensate 
in principle responds to chiral symmetry breaking/restoration.

The simulations have been carried out with a CUDA code in order to perform 
the simulations using the Hybrid Monte Carlo algorithm on GPU's. 

The parameters of our lattice calculation are the inverse of the bare 
coupling constant $\beta$, the bare mass $ma$ and the bare chiral chemical 
potential $a\mu_5$ (both in dimensionless units) as well as the lattice size. 
The physical temperature and the volume are given by
\begin{equation}\begin{split}
V &=(N_{\sigma} a(\beta))^3 \\
T &= \frac{1}{a(\beta)N_{\tau}}.
\end{split}\end{equation}

\begin{figure*}[h!]
\begin{tabular}{cc}
\includegraphics[scale=0.60,clip=false]{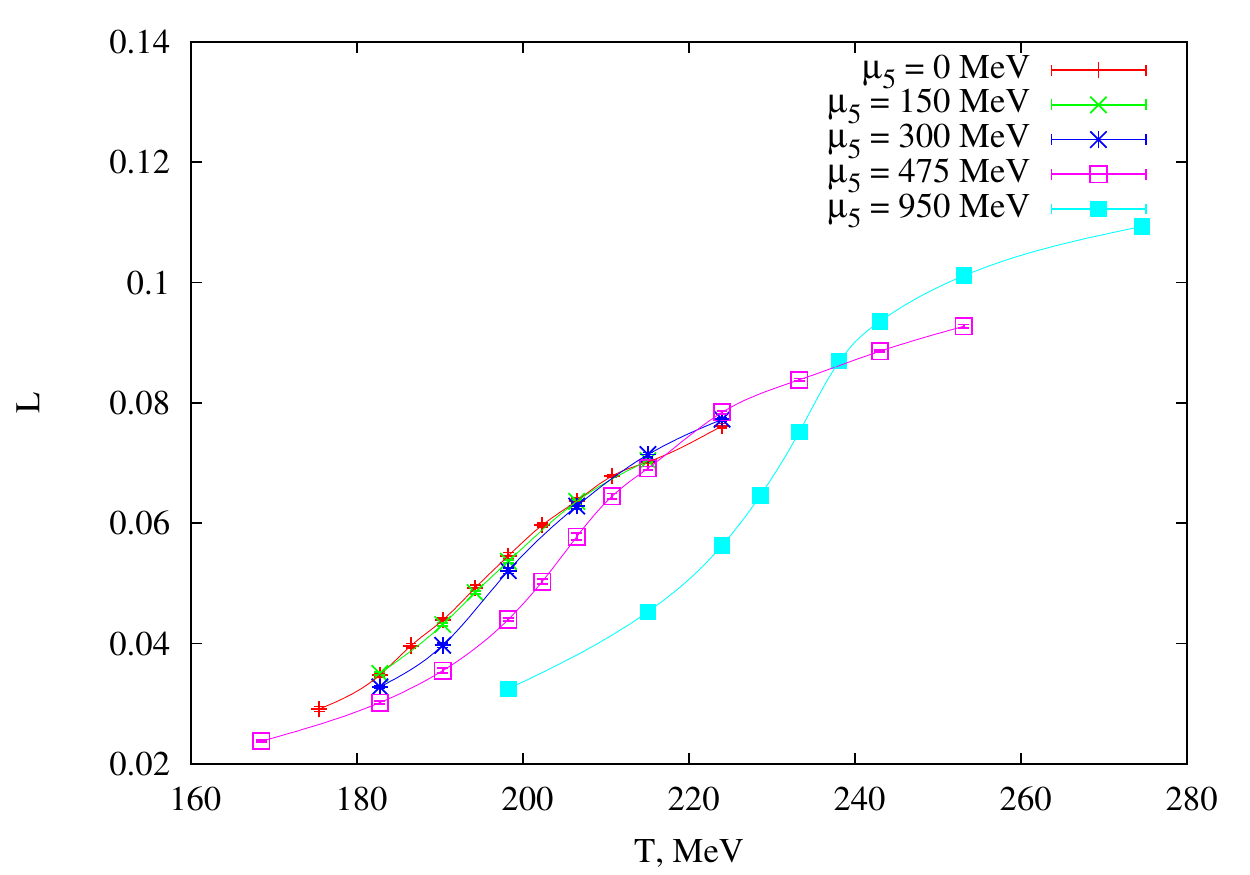} & 
\includegraphics[scale=0.60,clip=false]{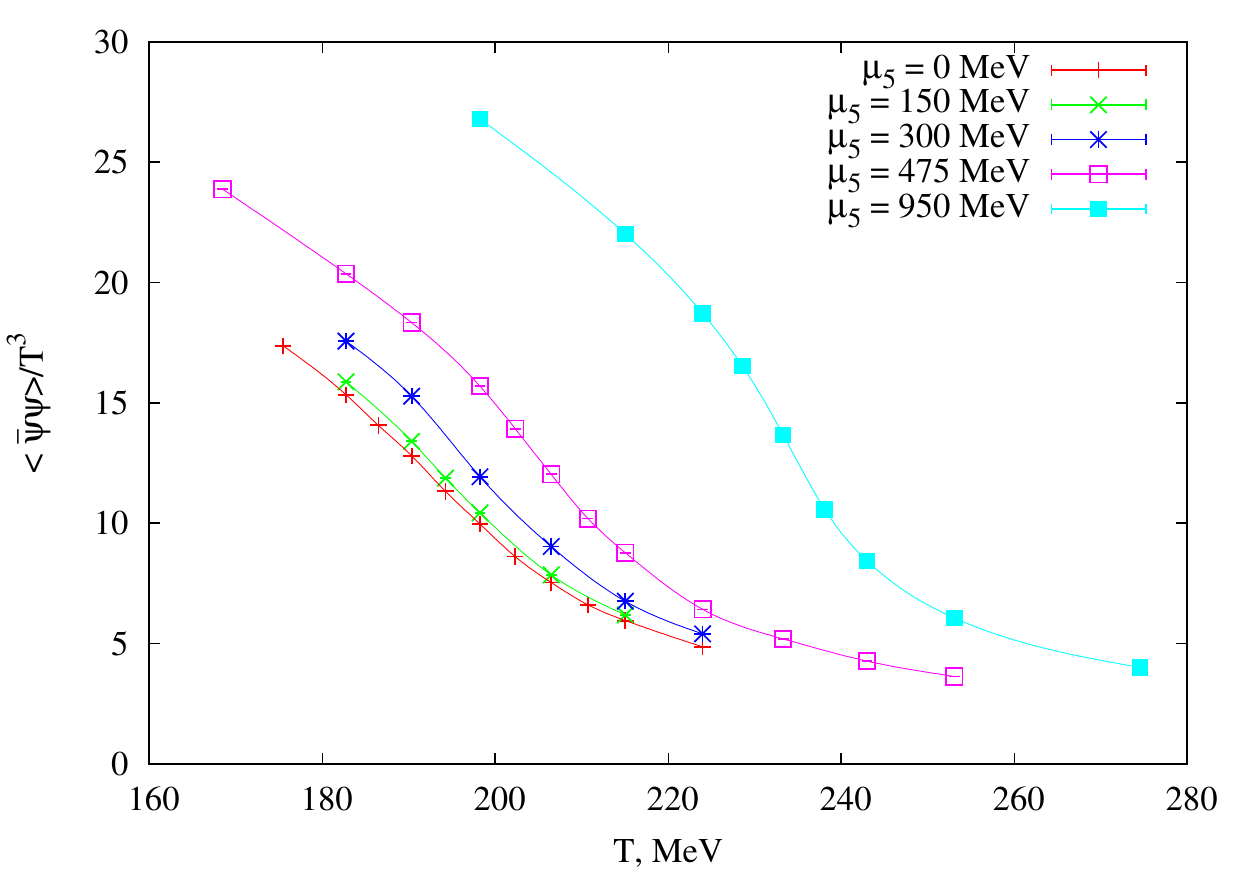} \\
Polyakov loop & chiral condensate
\end{tabular}
\caption{Polyakov loop and chiral condensate versus 
$T$ for five values of $\mu_5$. Lattice size is $6 \times 20^3$, fermion mass is $m\approx12$ MeV. Errors are smaller than the data point symbols. 
The curves are to guide the eyes.
}
\label{fig:small}
\end{figure*}

To perform the scale setting (calibration) of the lattice,
we use the results of Ref.~\cite{Ilgenfritz:2012fw}. There, the static 
potential in the same theory 
has been measured at zero temperature for several values of $\beta$. 
The Sommer parameter $r_0$ was calculated in lattice units and compared to 
its physical value, which was supposed to be the same as in QCD, i.e. 
$r_0=0.468(4)$ fm.  It was found that the scaling function $a(\beta)$ in 
the region were we perform our measurements is well described by the two 
loop $\beta$-function. Thus, for given $\beta$ we can obtain e.g. the 
temperature $T$ in units of MeV. For more details, see 
Ref.~\cite{Ilgenfritz:2012fw}. 
 
\section{Results of the calculation}

We first performed simulations on a lattice of size $6\times20^3$  for 
five fixed values of $\mu_5=0, 150, 300, 475, 950$ MeV and different 
values of $T$.
The fermion mass was kept fixed in physical units  
$m\approx 12$ MeV($m_{\pi}\approx 330$ MeV).
The expectation values of the Polyakov loop and the chiral condensate 
are shown in Fig. \ref{fig:small}. The sharp change of the observables 
as functions of $T$ indicates the onset of the deconfinement and
the chiral restoration phase transition. It is seen that the temperature 
of both phase transitions increases with the chiral chemical potential. 
One also sees that the phase trasition becomes sharper for increasing 
chiral chemical potential.

\begin{figure*}[h]
\begin{tabular}{cc}
\includegraphics[scale=0.60,clip=false]{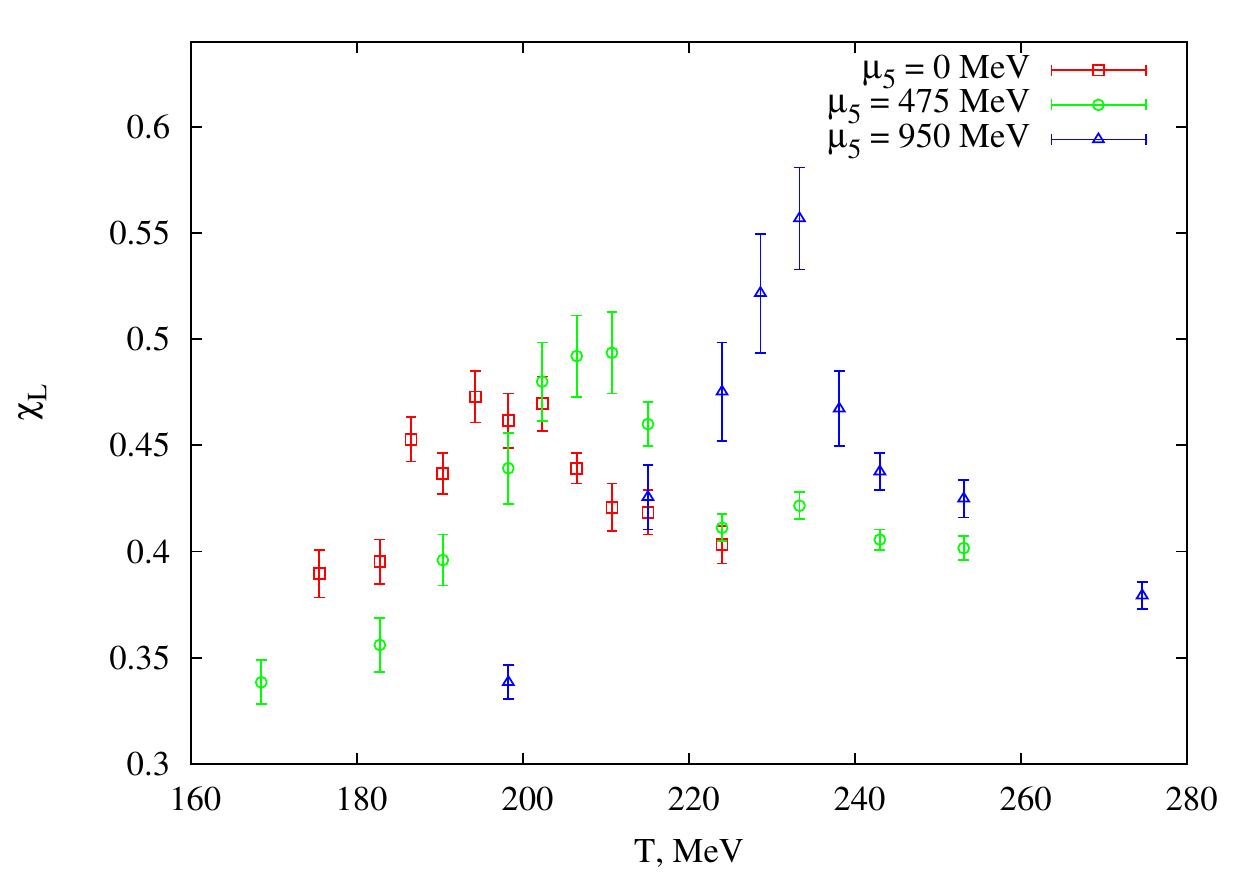} & 
\includegraphics[scale=0.60,clip=false]{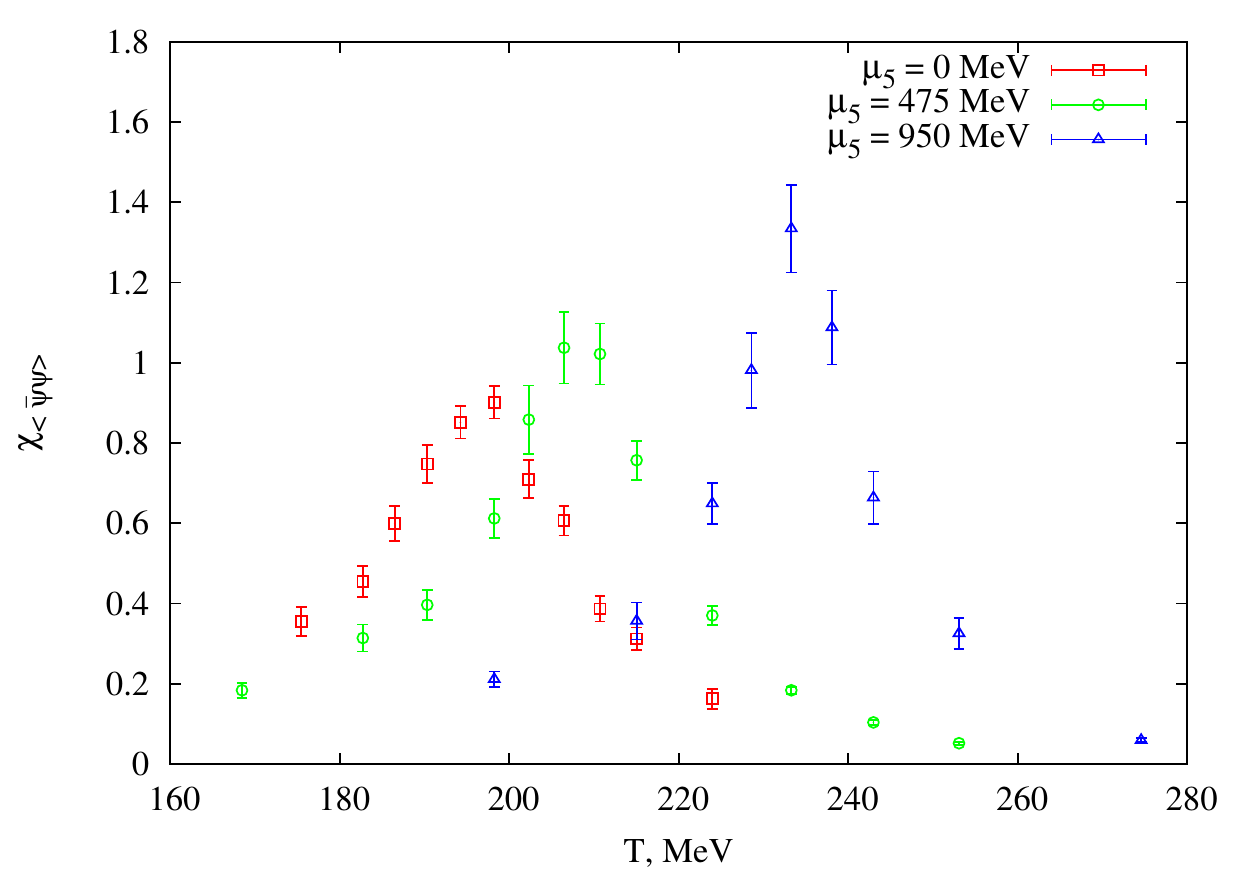} \\
Polyakov loop susceptibility & chiral susceptibility
\end{tabular}
\caption{Polyakov loop susceptibility and chiral susceptibility versus 
$T$ for three values of $\mu_5=0, 475, 950$ MeV. Lattice size is $6 \times 20^3$, fermion mass is $m\approx12$ MeV. 
In order to avoid a complete superposition of data points belonging to 
different $\mu_5$ values we applied a tiny shift along the $T$ axis.}
\label{fig:smallsus1}
\end{figure*}

\begin{figure*}[t]
\begin{tabular}{cc}
\includegraphics[scale=0.60,clip=false]{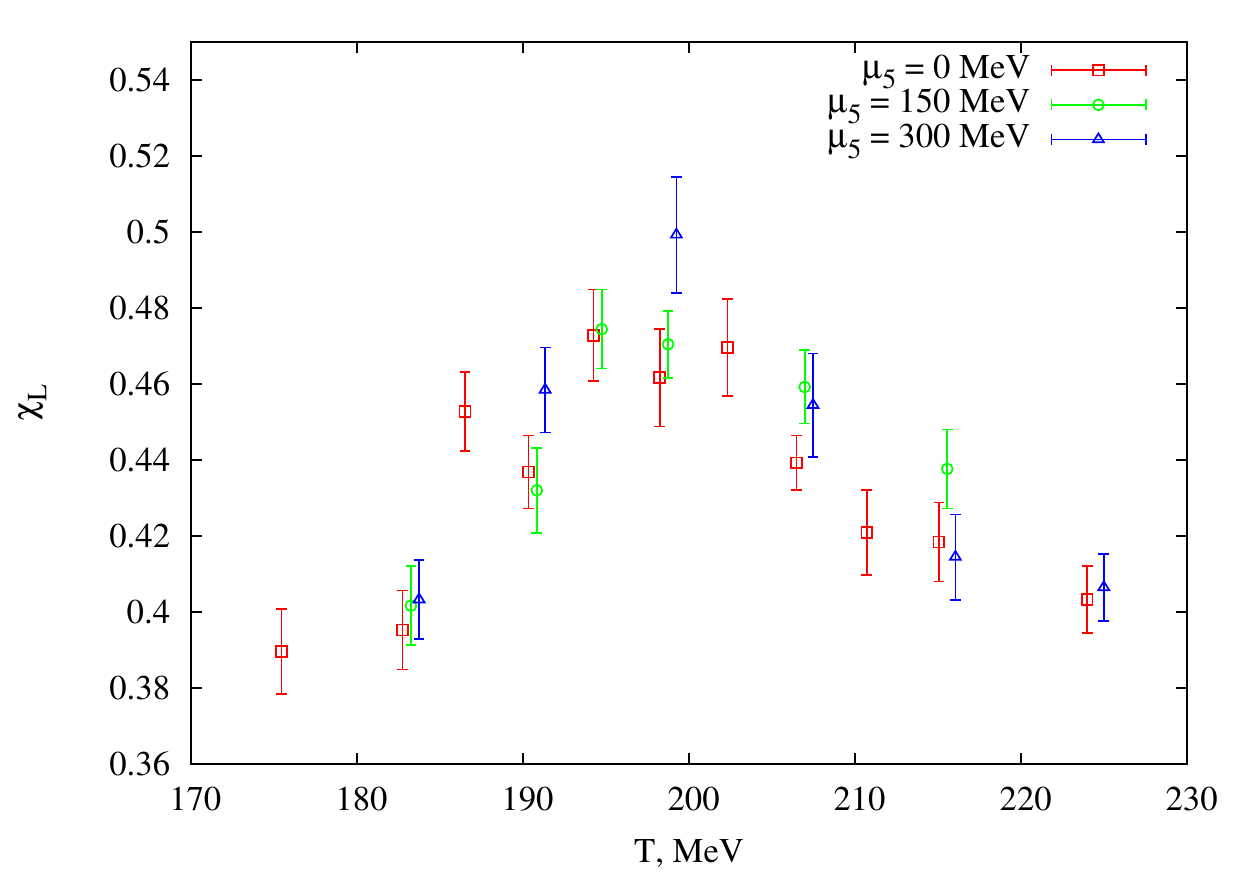} & 
\includegraphics[scale=0.60,clip=false]{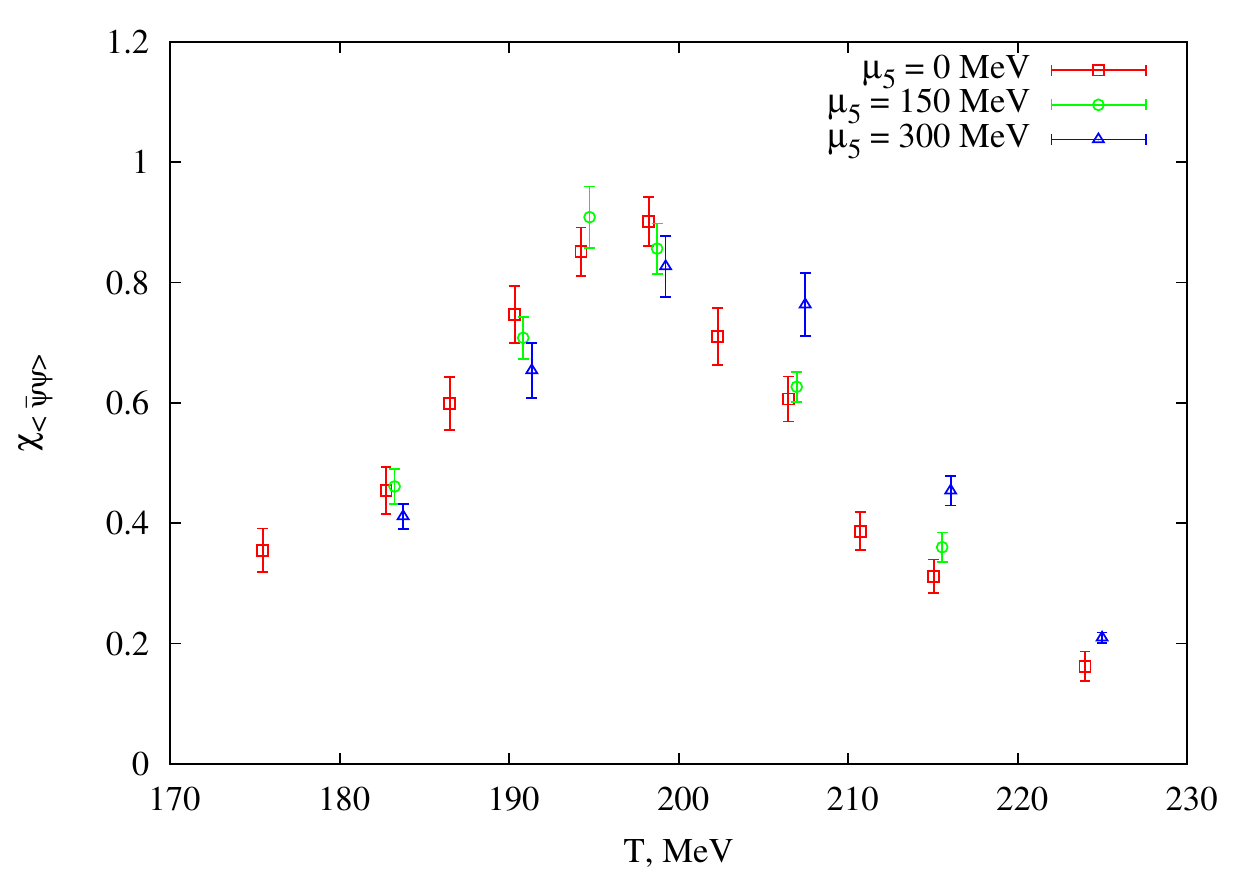} \\
Polyakov loop susceptibility & chiral susceptibility
\end{tabular}
\caption{Polyakov loop susceptibility and chiral susceptibility versus 
$\beta$ for three values of $\mu_5=0, 150, 300$ MeV. Lattice size is $6 \times 20^3$, fermion mass is $m\approx12$ MeV. 
In order to avoid a complete superposition of data points belonging to 
different $\mu_5$ values we applied a tiny shift along the $T$ axis.}
\label{fig:smallsus2}
\end{figure*}

\begin{figure*}[h]
\includegraphics[scale=0.60,clip=false]{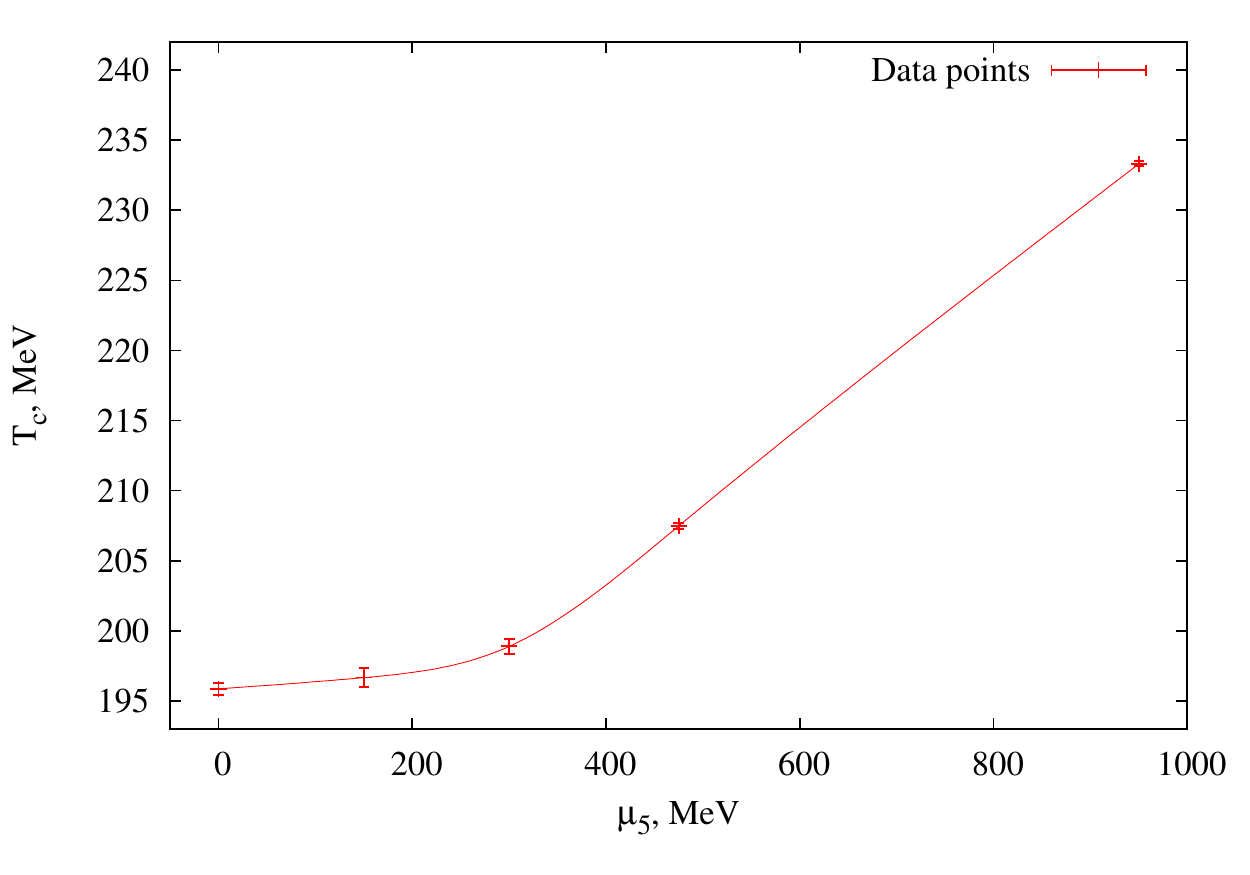} 
\caption{The dependence of the critical temperature on the value of the chiral chemical potential. Lattice size is $6 \times 20^3$, fermion mass is $m\approx12$ MeV. The curve is to guide the eyes.}
\label{fig:crittemp}
\end{figure*}

To study the change of the critical temperature more quantitatively, 
we also calculated the chiral and the Polyakov loop susceptibilities. 
In order to make the figures readable we plot separately the susceptibilities 
for values $\mu_5=0,475,950$ MeV in Fig. \ref{fig:smallsus1} and for values
$\mu_5=0, 150,300$ MeV in Fig. \ref{fig:smallsus2}.
We see that increasing the value of the chiral chemical potential moves the 
position of the peaks of the chiral and Polyakov loop susceptibilities to 
larger values of $\beta$. This means that the transition temperature increases.
Our results do not show any splitting between the chiral and the deconfinement 
transition. We have fitted the data for the chiral susceptibility near the 
peak with a gaussian function 
$\chi=a_0+a_1\exp{\left(-\frac{(\beta-\beta_c)^2}{2\sigma^2}\right)}$ 
and extracted the critical temperature $T_c(\beta_c)$. The resulting 
dependence of the critical temperature on the value of the chiral chemical 
potential $\mu_5$ is shown on the Fig. \ref{fig:crittemp} and 
Table \ref{tab:ss}. The values of  $\beta_c$ and $T_c$ and their uncertainties are
calculated from a fit of 5-6 points in the vicinity of the peak by the
Gaussian function given above. It should be noted that for small values of $\mu_5$ 
the dependence of $T_c$ on $\mu_5$ is well described by the function 
$T_c=a+b \mu_5^2$, but the larger $\mu_5$ is the larger is the deviation 
from this simple formula.

\begin{table*}[h]
\mbox{
\begin{tabular}{|c|c|c|}
\hline
$\mu_5$ (MeV) & $\beta_c$ & $T_c$ (MeV)  \\
\hline
0   & 1.7975 $\pm$ 0.0005 & 195.8 $\pm$ 0.4\\
150 & 1.7984 $\pm$ 0.0009 & 196.7 $\pm$ 0.7 \\
300 & 1.8012 $\pm$ 0.0007 & 198.9 $\pm$ 0.5 \\
475 & 1.8116 $\pm$ 0.0003 & 207.5 $\pm$ 0.2 \\
950 & 1.8404 $\pm$ 0.0002 & 233.3 $\pm$ 0.2 \\
\hline
\end{tabular}
}
\caption{
The critical temperature $T_c$ and the lattice parameter$\beta_c$ as a 
function of the chiral chemical potential $\mu_5$ obtained from the fit 
of the chiral susceptibility near the peak by a gaussian function 
$\chi=a_0+a_1\exp{\left(-\frac{(\beta-\beta_c)^2}{2\sigma^2}\right)}$. }
\label{tab:ss}
\end{table*}

In addition to the calculations on the $6 \times 20^3$ lattice we carried 
out simulations on a larger lattice of size $10 \times 28^3$. Note that
this allows us to investigate larger values of $\mu_5$. The susceptibilities 
require large statistics, and therefore our current computational resources 
do not allow us to measure them on the larger lattice. In the simulations
we kept the physical fermion mass fixed at 
$m=18.5$ MeV ($m_{\pi} \approx 550$ MeV).

\begin{figure*}[t]
\begin{tabular}{cc}
\includegraphics[scale=0.60,clip=false]{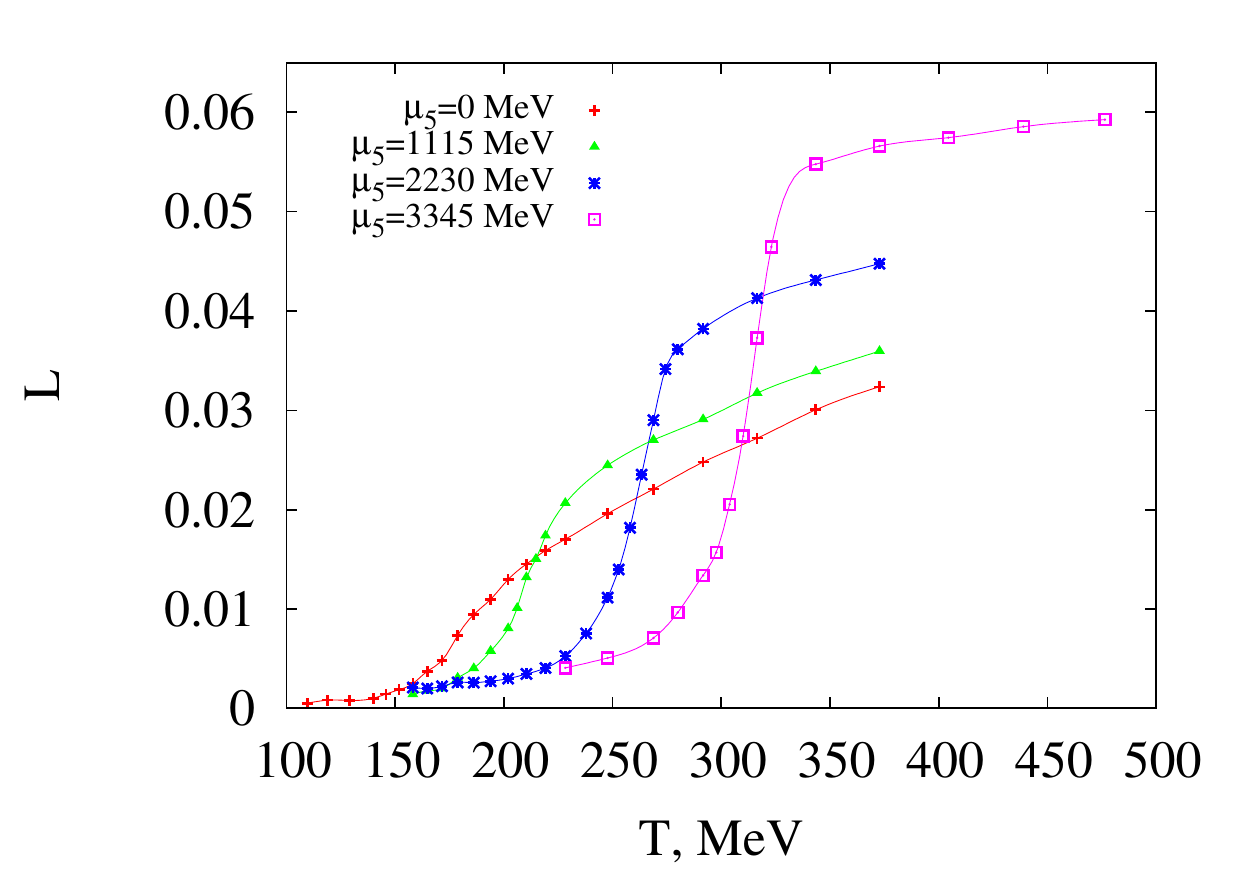} & 
\includegraphics[scale=0.60,clip=false]{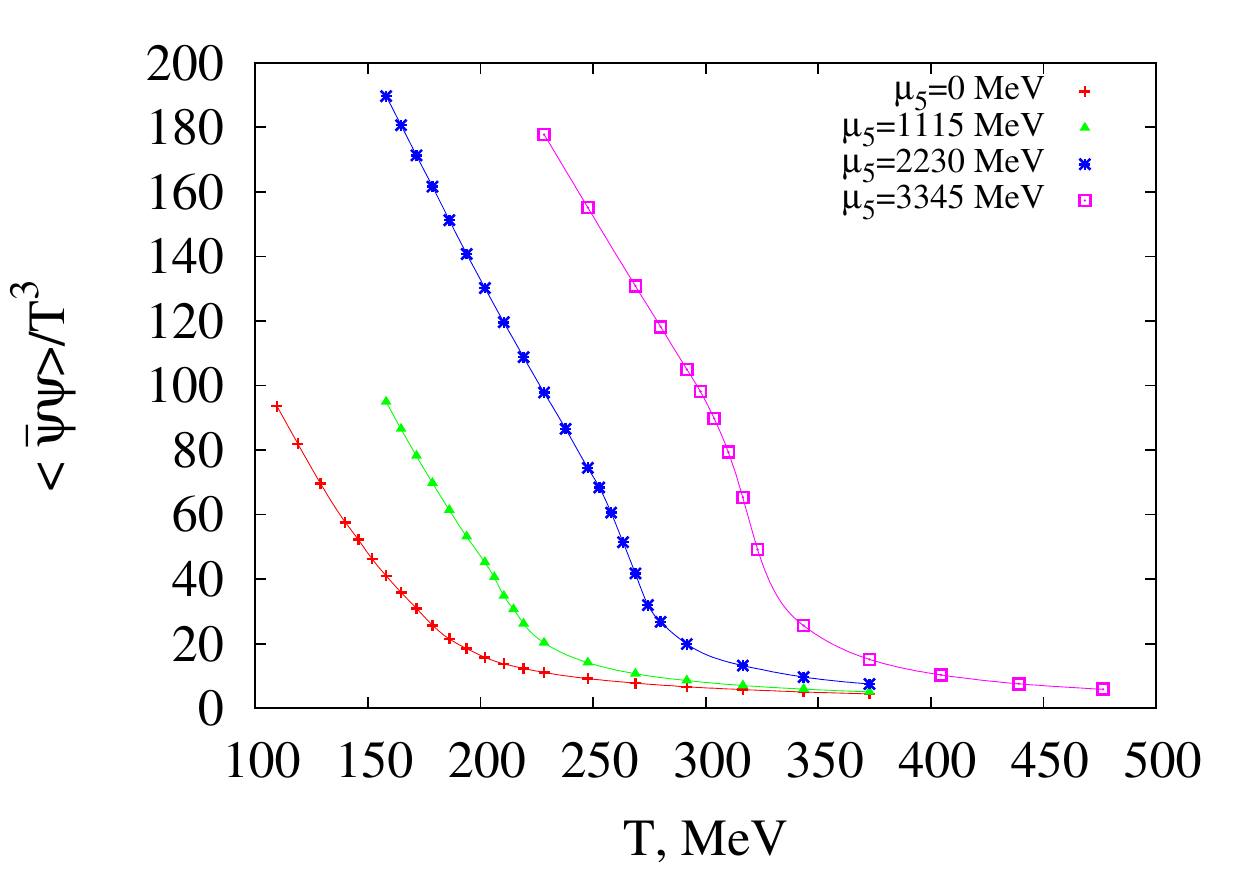} \\
Polyakov loop & chiral condensate
\end{tabular}
\caption{Polyakov loop and chiral condensate versus $T$ for four values of $\mu_5$  and lattice size $10\times28^3$, fermion mass is $ma\approx18.5$ MeV. 
Errors are smaller than the data point symbols. The curves are to guide the eyes.}
\label{fig:large}
\end{figure*}

On this lattice we also calculated the observables $L$ and 
$\langle \psi \psi \rangle$ as functions of $T$ for different values of 
$\mu_5$. The results for the Polyakov loop and the chiral condensate are 
presented in Fig. \ref{fig:large}.
They underpin the fact that the critical temperature grows when 
one increases $\mu_5$. The value of the chiral chemical potential used in 
our simulations was rather large (up to $\mu_5=3345$ MeV), but we have not 
seen signals of a first-order phase transition, although the transition 
becomes sharper when the value of the chiral chemical potential grows.

\begin{figure*}[h]
\begin{tabular}{cc}
\includegraphics[scale=0.60,clip=false]{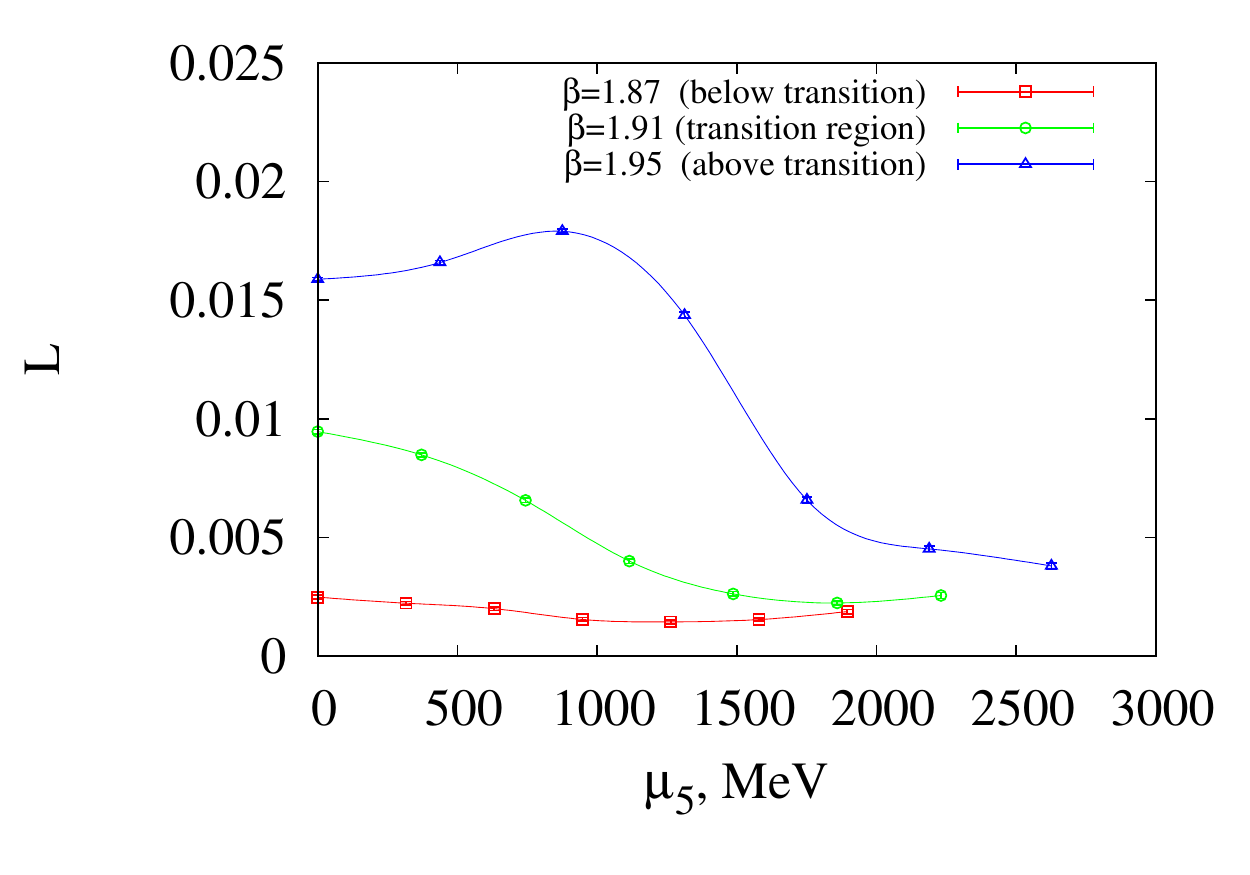} & 
\includegraphics[scale=0.60,clip=false]{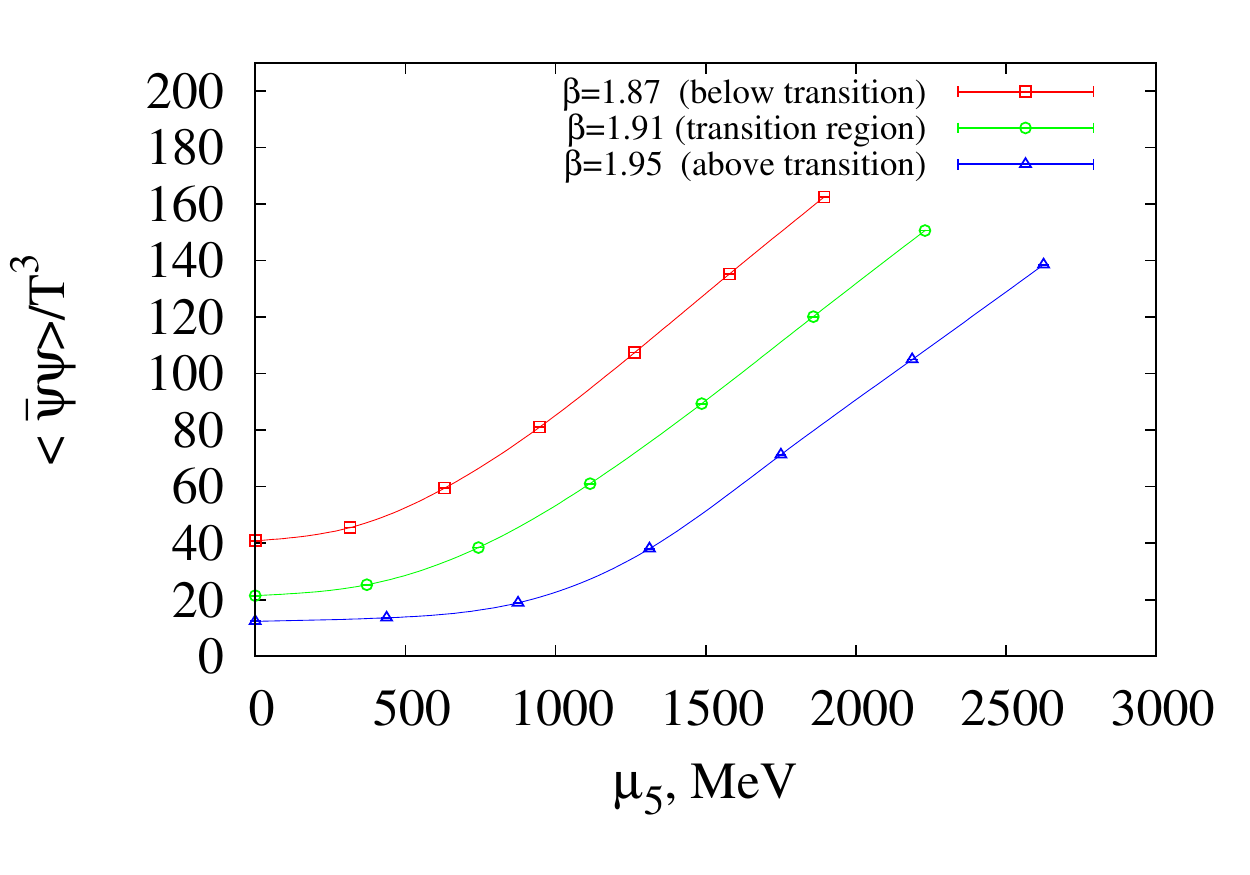} \\
Polyakov loop & chiral condensate
\end{tabular}
\caption{Polyakov loop and chiral condensate versus $\mu_5$ for three $\beta$ values and lattice size $10\times28^3$, fermion mass is $ma\approx18.5$ MeV. 
Errors are smaller than the data point symbols. The curves are to guide the eyes.}
\label{fig:largeconst}
\end{figure*}

It is interesting to study how the observables $L$ and 
$\langle \psi \psi \rangle$ depend on $\mu_5$ for fixed temperature.  
To do this we have measured the Polyakov loop and the chiral condensate 
as functions of $\mu_5a$ for three different values of $\beta$,
$\beta=1.87$ (158 MeV), $1.91$ (186 MeV) and $1.95$ (219 MeV), which for 
a vanishing chiral chemical potential corresponds to a temperature below 
the transition, in the transition region, and in the high temperature phase, 
respectively. The results of these measurements are shown in 
Fig. \ref{fig:largeconst}. As can be seen from the figure, in the confinement 
phase the Polyakov loop remains almost constant with increasing chiral 
chemical potential. It means that if the system was in the confinement 
phase at $\mu_5 = 0$, it remains confined at $\mu_5 > 0$.
Moreover, we observe the Polyakov loop to drop down with increasing $\mu_5$
both in the deconfinement phase and in the transition region.
Thus, the system goes into the confinement phase for sufficiently large 
$\mu_5$. With other words, we conclude that the critical temperature 
increases with an increasing chiral chemical potential in agreement with 
our results obtained on the smaller lattice. It is worth mentioning that 
the behavior described above looks quite similar to the behavior obtained 
for two-color QCD in an external magnetic 
field\cite{Ilgenfritz:2012fw, Ilgenfritz:2013oda}.

\begin{figure*}[tb]
\begin{tabular}{cc}
\includegraphics[scale=0.60,clip=false]{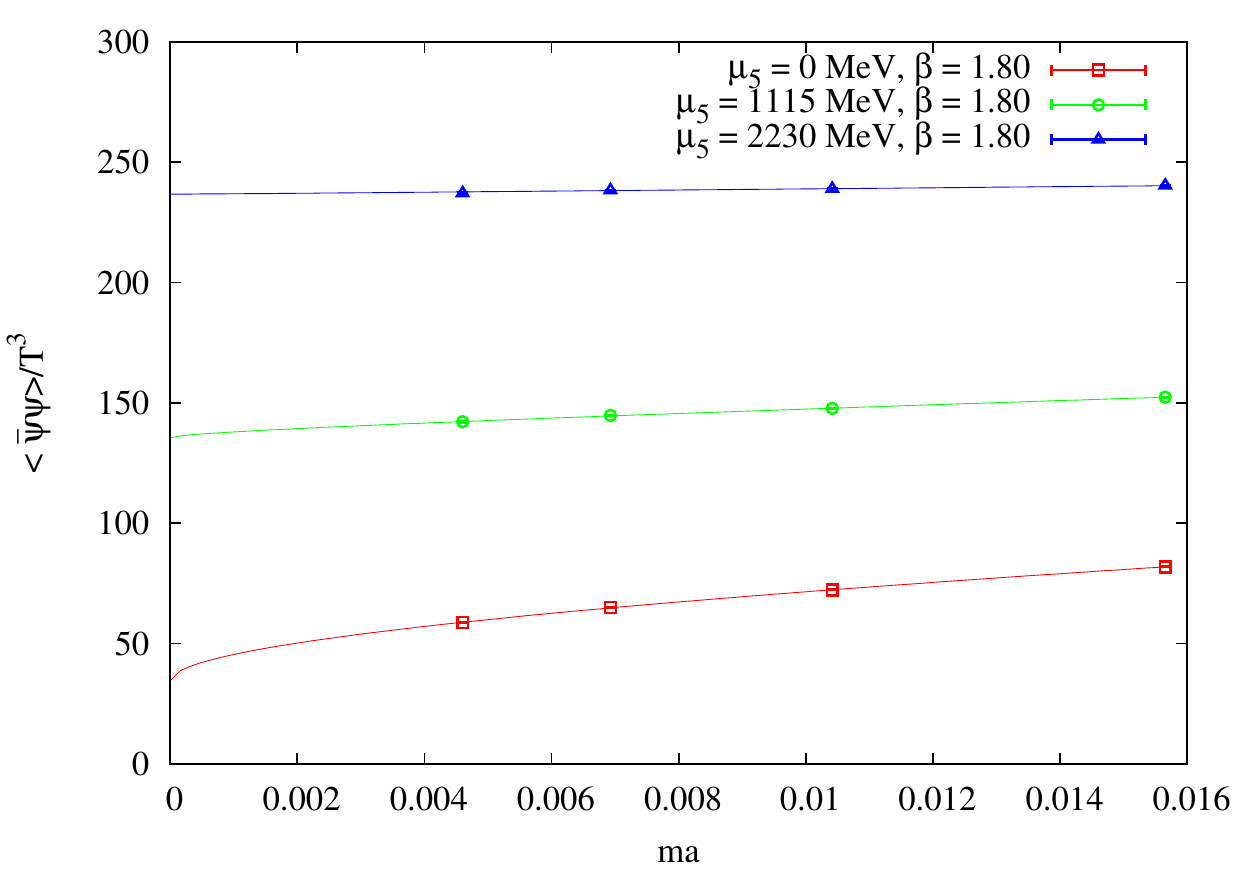} & 
\includegraphics[scale=0.60,clip=false]{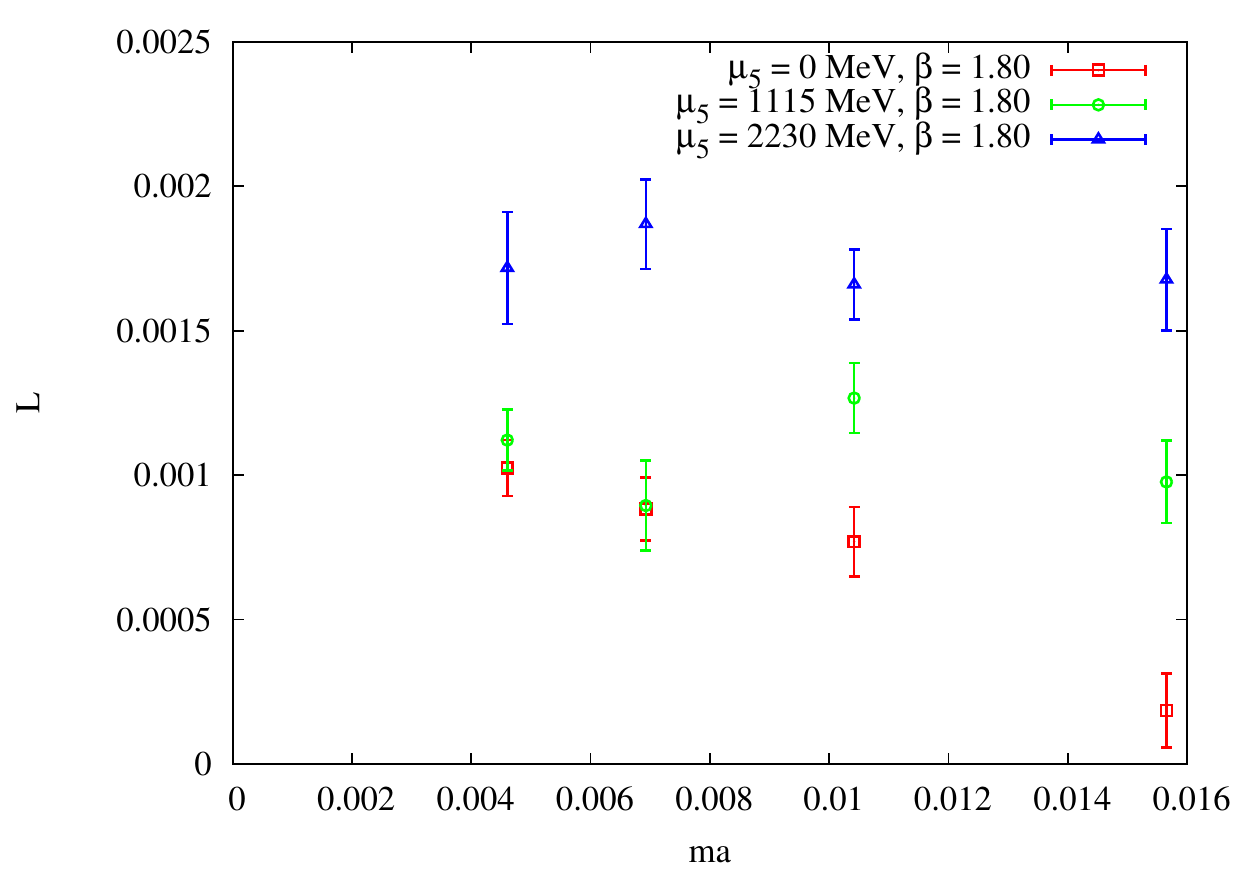} 
\end{tabular}
\caption{Chiral condensate and Polyakov loop versus $ma$ for
three $\mu_5$ values at $\beta=1.80$ on a lattice of size
$10 \times 28^3$. $\beta=1.80$ corresponds to the confinement phase.}
\label{fig:chiral180}
\end{figure*} 

\begin{figure*}[b]
\begin{tabular}{cc}
\includegraphics[scale=0.60,clip=false]{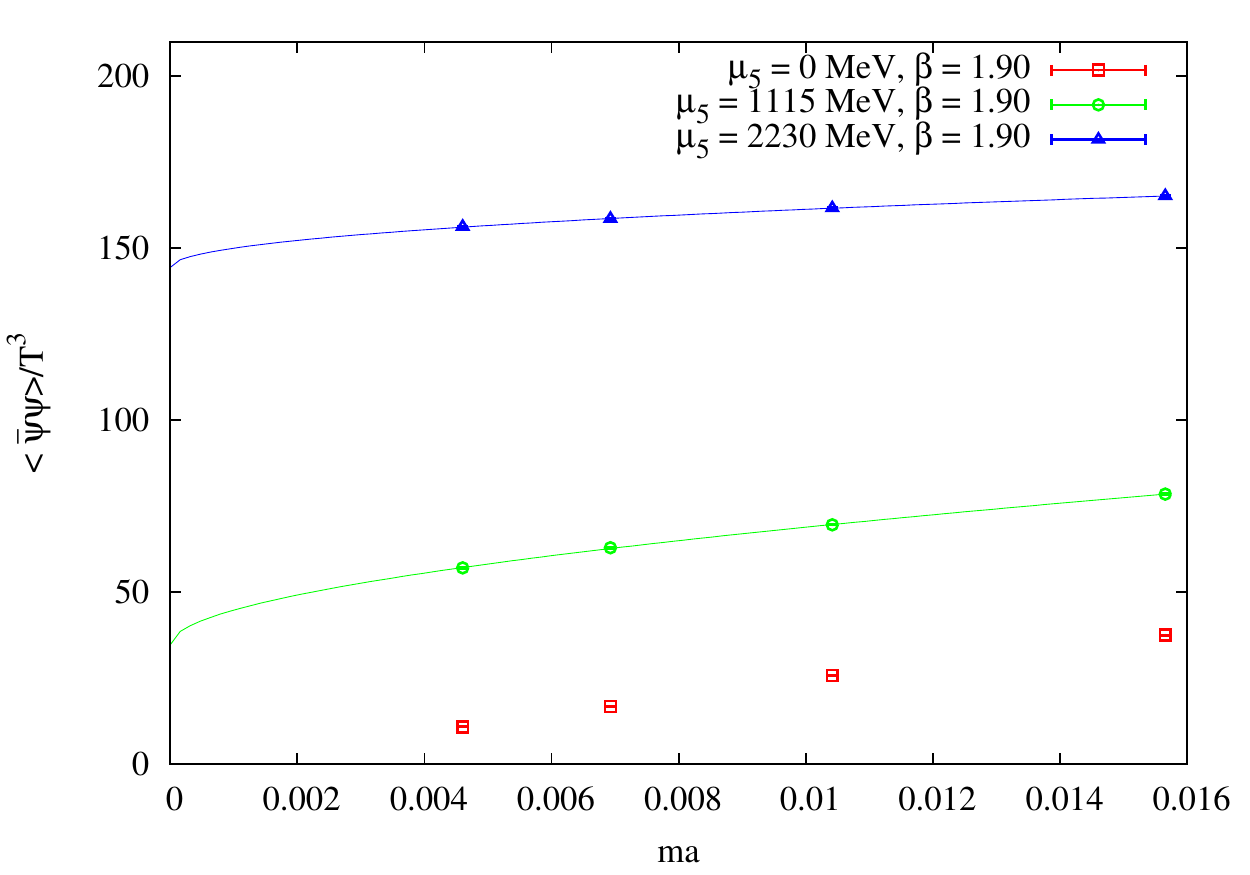} & 
\includegraphics[scale=0.60,clip=false]{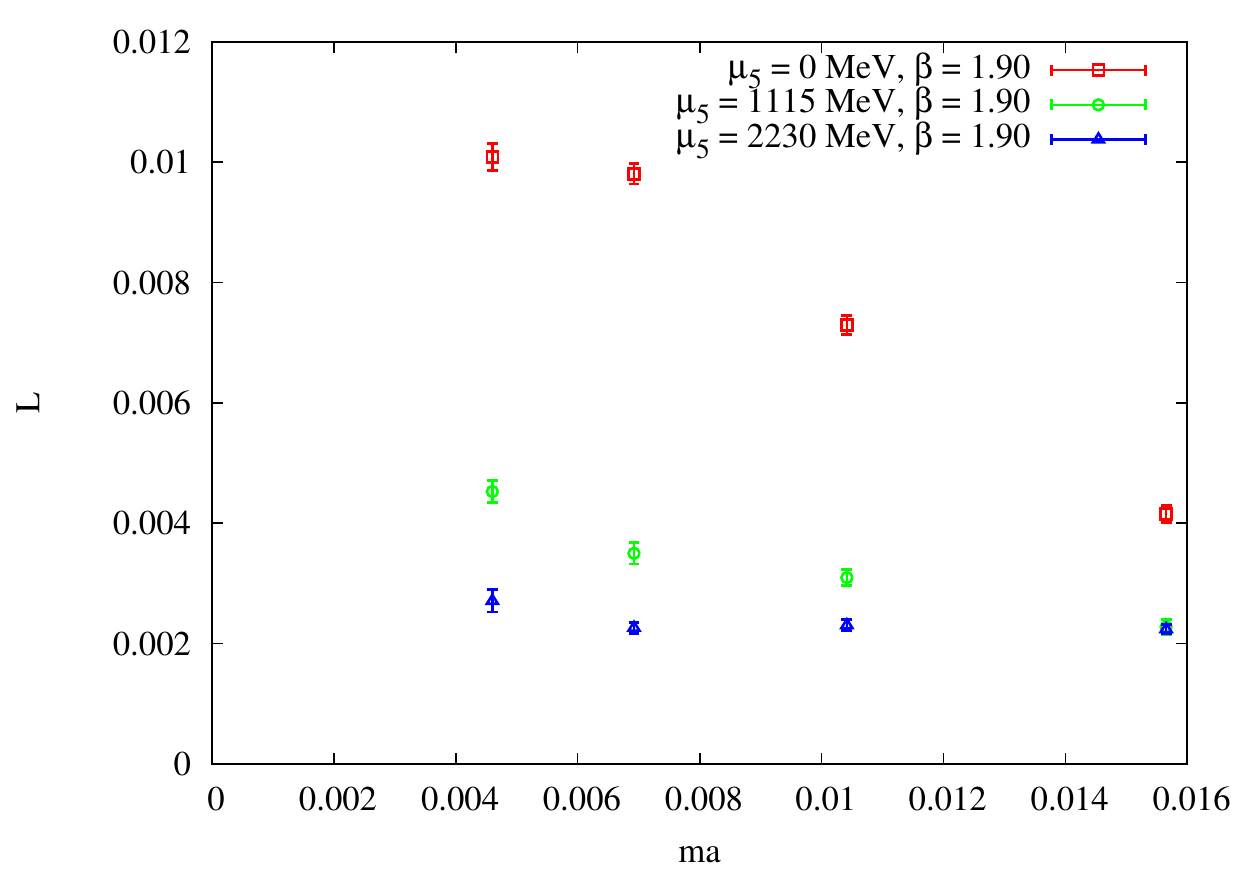} 
\end{tabular}
\caption{Same as Fig. 7, but for $\beta=1.90$ corresponding to the
transition region.}
\label{fig:chiral190}
\end{figure*}

\begin{figure*}[t]
\begin{tabular}{cc}
\includegraphics[scale=0.60,clip=false]{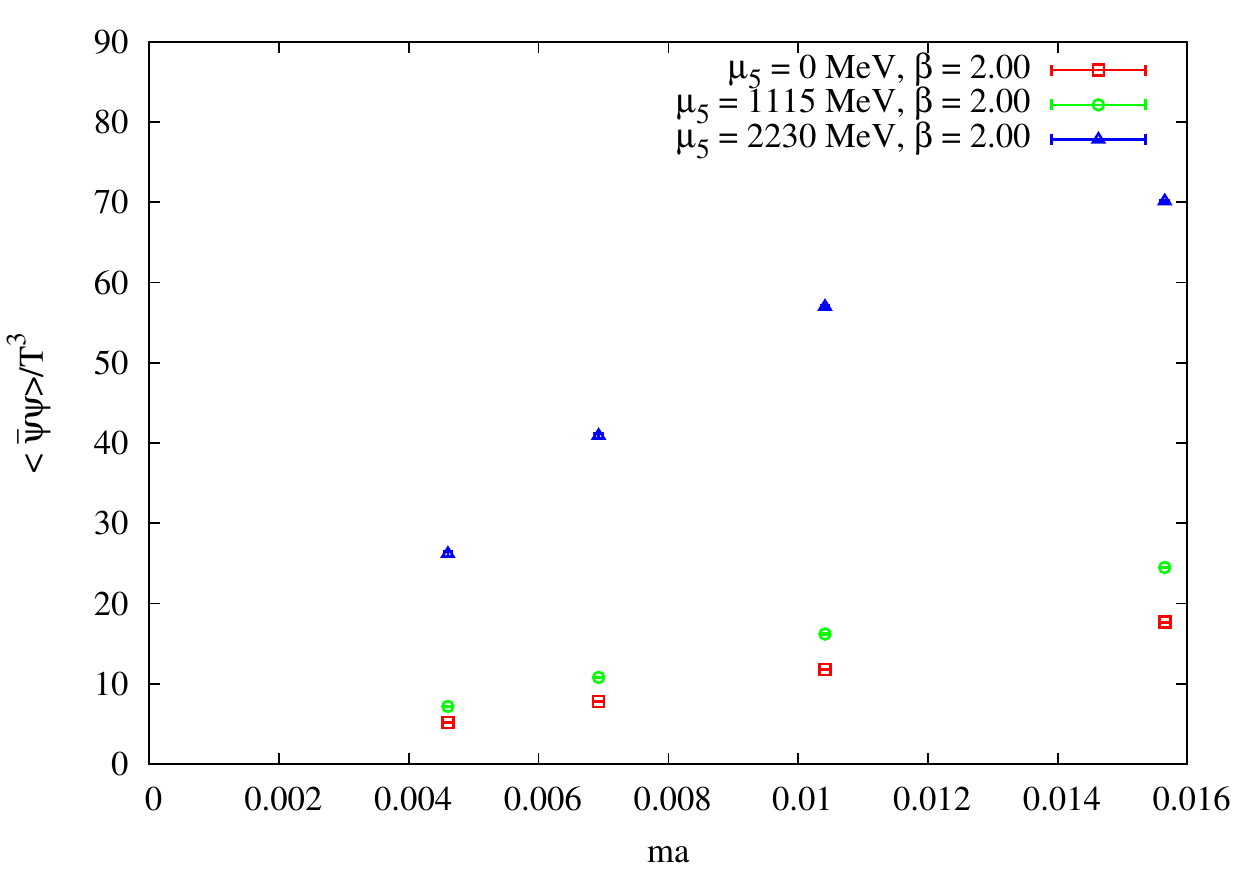} & 
\includegraphics[scale=0.60,clip=false]{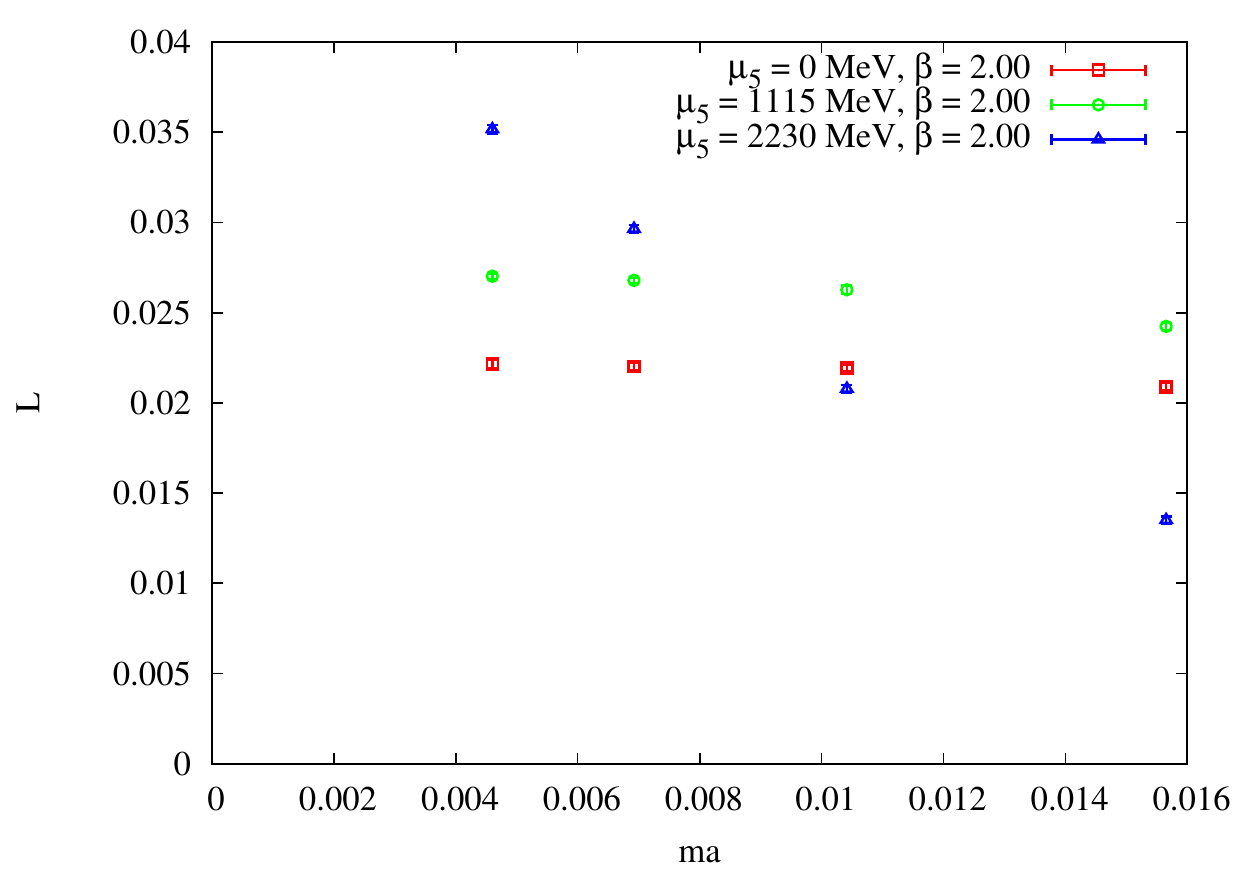} 
\end{tabular}
\caption{Same as Fig. 7, but for $\beta=2.00$ corresponding to the
transition region as well..}
\label{fig:chiral200}
\end{figure*}

\begin{figure*}[t]
\begin{tabular}{cc}
\includegraphics[scale=0.60,clip=false]{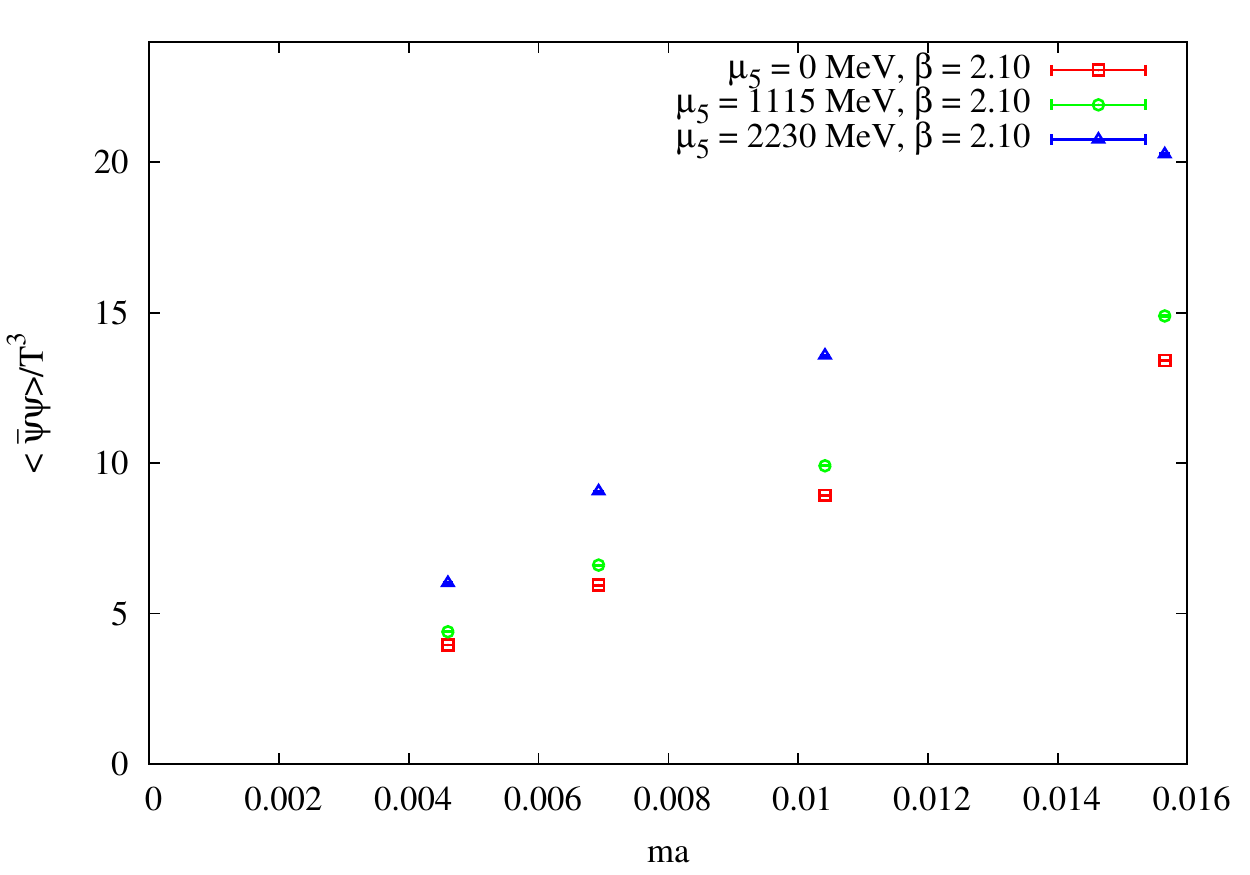} & 
\includegraphics[scale=0.60,clip=false]{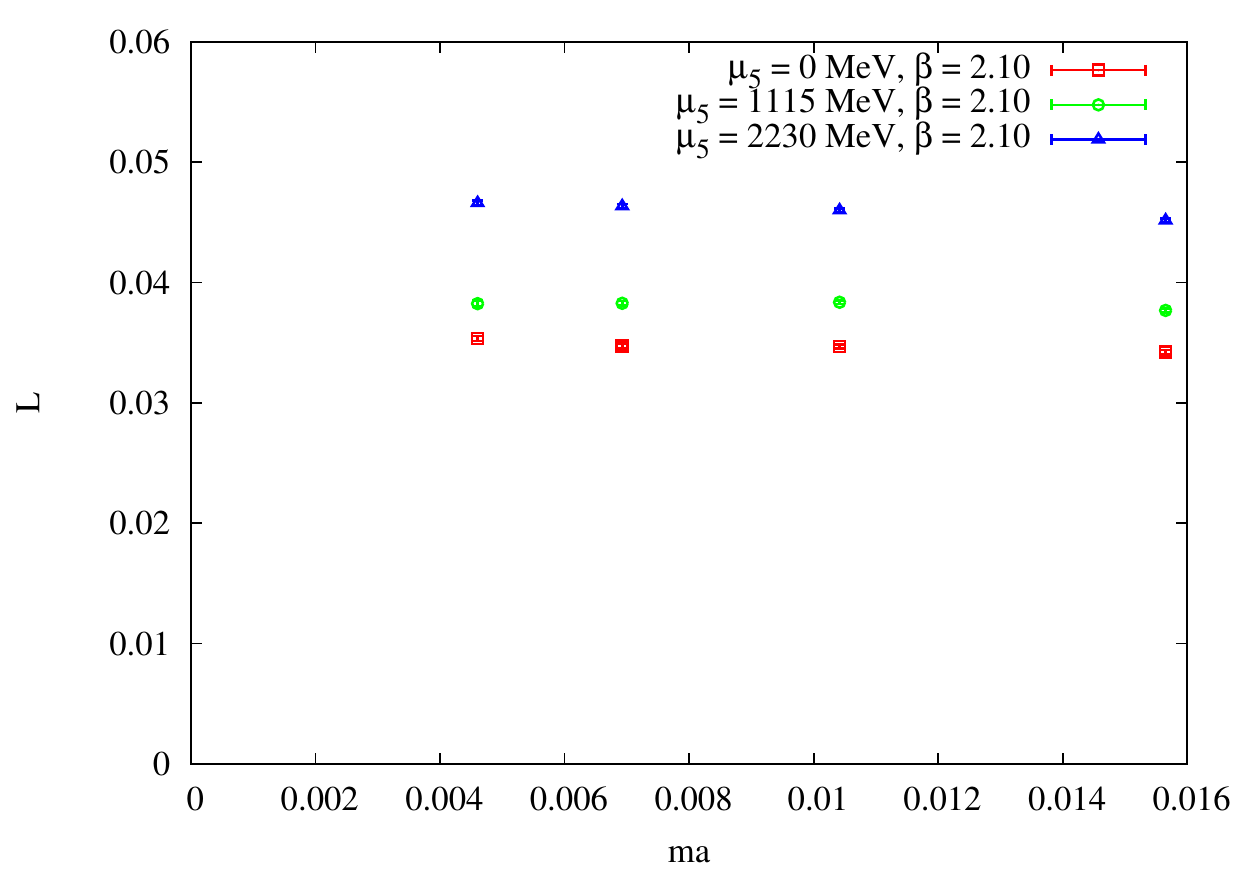} 
\end{tabular}
\caption{Same as Fig. 7, but for $\beta=2.10$ corresponding to the
deconfinement phase.}
\label{fig:chiral210}
\end{figure*}

In order to study the dependence of the observables on the value of the 
quark mass for different values of chiral chemical potential, we chose 
four values of $\beta$: one in the confinement phase, $\beta=1.8$ 
(corresponding to a temperature $T=119$ MeV), 
two in the transition region, $\beta=1.9$ ($T=178$ MeV) and $\beta=2.0$ 
($T=268$ MeV), and one in the deconfinement phase $\beta=2.1$ ($T=404$ MeV). 
The lattice size used here is $10\times28^3$. We chose four values of the 
fermion mass $ma$ and three values of the chemical potential
 $\mu_5=0$ MeV, $\mu_5=1115$ MeV and $\mu_5=2230$ MeV. The expectation 
values of the Polyakov loop and the chiral condensate are presented as 
functions of $ma$ for different $\beta$ in Fig. \ref{fig:chiral180}, 
Fig. \ref{fig:chiral190}, Fig. \ref{fig:chiral200}, and
Fig. \ref{fig:chiral210}.

\begin{table*}[h]
\mbox{
\setlength{\tabcolsep}{1.0pt}
\begin{tabular}{|c|c|c|c|c|c|c|c|c|c|}
\hline
 \multicolumn{2}{|c|}{}
 & \multicolumn{4}{|c|}{$f_1(ma)$}
 & \multicolumn{4}{|c|}{$f_2(ma)$} \\
\hline
$\beta$ & $\mu_5$, MeV & $\chi^2_{dof}$ & $a_0$ & $a_1$ &  $a_2$ & $\chi^2_{dof}$ & $b_0$ & $b_1$ & $b_2$ \\
\hline
1.80 & 0 & 0.56 & 0.034(1) & 0.34(2) & 0.3(1) &  1.2 & 0.042(1) & -0.9(1) & -1.1(4)\\
1.80 & 1115 & 1.83 & 0.135(2) & 0.06(5) & 0.6(2) &
2.0 & 0.137(1) & -0.15(13) & 0.4(5)\\
1.80 & 2230 & 2.0 & 0.237(5) & -0.004(100) & 0.3(5) & 2.0 & 0.237(3) & 0.02(26) & 0.3(9)\\
1.90 & 1115 & 2.7 & 0.034(3) & 0.31(6) & 0.3(3) & 3.9 & 0.042(2) &  -0.8(2) & -1.0(7)\\
1.90 & 2230 & 0.62 & 0.144(2) & 0.18(3) & -0.11(15) & 0.3 & 0.1486(5) & -0.47(6) & -0.9(2)\\
\hline
\end{tabular}
}
\caption{The parameters for the fits $f_1(ma)$ Eq. (\ref{eq:3d}) and  $f_2(ma)$ Eq. (\ref{eq:4d})
allowing to extrapolate to the chiral limit for $\beta=1.80$ and $\beta=1.90$  and
various values of chiral chemical potential.
The fit curves obtained with $f_1$ are shown in the left panel of Fig. \ref{fig:chiral180}, \ref{fig:chiral190}.}
\label{tab:tab1}
\end{table*}

In Fig. \ref{fig:chiral180} we show the results in the confinement region
at $\beta = 1.80$ ($T=119$ MeV). The Polyakov loop is small and does not
show any nontrivial behaviour. The chiral condensate remains almost constant 
when the value of the fermion mass changes.

To extrapolate to the chiral limit $(ma=0)$ we performed chiral extrapolations 
in two ways. Since the value of the temperature is not very far from the 
transition temperature, we suppose that we can use the formula for the 
reduced three dimensional model \cite{Pisarski:1983ms}.
See also \cite{Wallace:1975vi,Hasenfratz:1989pk,Smilga:1993in}. 
In that case we may use the ansatz $a^3<\bar{\psi}\psi>=f_1(ma)$, where
\begin{equation}\begin{split}
 f_1(ma) = a_0 + a_1 \sqrt{ma} + a_2 ma 
\label{eq:3d}
\end{split}\end{equation}
with the non-analytic term coming from the Goldstone bosons.
Such a parametrization has been used in \cite{Bazavov:2011nk}
in the context of the finite temperature transition in QCD.
As an alternative we also use the chiral extrapolation
relevant for zero temperature, namely
$a^3<\bar{\psi}\psi>=f_2(ma)$, where
\begin{equation}\begin{split}
 f_2(ma) = b_0 + b_1 ma \log ma + b_2 ma 
\label{eq:4d}.
\end{split}\end{equation}

The fits performed with the Eq. (\ref{eq:3d}) are shown as lines in 
Fig. \ref{fig:chiral180}. The corresponding parameters of both are
given in Table \ref{tab:tab1}.
As can be seen from table II and the corresponding figure, the chiral 
condensate extrapolates to a non-zero value in the chiral limit.

At larger values $\beta=1.90$ ($T=178$ MeV), $\beta=2.00$ ($T=268$ MeV), 
and $\beta=2.10$ ($T=404$ MeV)  and at zero chiral chemical potential 
the system is in the chirally restored phase, when $m$ goes to zero.
However, at $\beta=1.90$ ($T=178$ MeV) and $\mu_5=1115, 2230$ MeV, the 
chiral condensate has non-zero expectation values.
We have performed the same extrapolation technique as for $\beta=1.80$ 
and present also these results in Table \ref{tab:tab1}, showing that 
for these values of $\mu_5$ the data are consistent with a non-zero 
values in the chiral limit. 
It implies that also in the chiral limit a non-zero chiral chemical 
potential shifts the position of the phase transition to larger temperatures. 
The plot for the Polyakov loop (right panel of Fig. \ref{fig:chiral190})
confirms this observation: for the two smaller values of the fermion mass 
$ma$ the values of Polyakov loop for $\mu_5a=0$ is several times larger 
than for non-zero values of $\mu_5$, so that the system for non-zero 
$\mu_5$ is deeper in the confinement region than for zero chiral chemical 
potential.

At $\beta=2.00$ ($T=268$ MeV) and $\beta=2.10$ ($T=404$ MeV) the chiral 
condensate goes to zero in the  chiral limit (Fig. \ref{fig:chiral200},
 \ref{fig:chiral210}), thus the system is in the chirally restored phase. 
The dependence of the chiral condensate on the value of the mass is almost 
a linear function, apart from the behavior for $\beta=2.00$ ($T=268$ MeV) 
at the largest chiral chemical potential under our investigation, 
$\mu_5=2230$ MeV. This behaviour can be explained if one assumes that 
this point is near the phase transition. It is known that increasing the 
mass shifts the position of the transition to larger temperatures 
\cite{Ilgenfritz:2012fw}. Thus, at smaller masses the system is in the 
chirally restored phase, while increasing the mass lets the system undergo 
the chiral phase transition. Again, the plot of the Polyakov loop confirms 
this suggestion. The Polyakov loop is small at larger masses implying that 
the system is in the confinement phase and increases when the mass is 
decreased, i.e. the system becomes deconfining. This behavior suggests that 
at the largest value of $\mu_5=2230$ MeV the transition happens at even 
larger temperature than we have investigated. These results are in total
agreement with the results above.

\section{Discussion and conclusion}

In this paper we have presented an investigation of the phase diagram of 
two-color QCD with a chiral chemical potential using lattice simulations
with dynamical staggered fermions without rooting, i.e. four flavors in 
the continuum limit. We have calculated the chiral condensate, the Polyakov 
loop, the Polyakov loop susceptibility and the chiral susceptibility for 
different values of the temperature $T$ and the chiral chemical potential 
$\mu_5$ on lattices of size $6 \times 20^3$ and $10 \times 28^3$. The main 
result is that at non-zero values of the chiral chemical potential the 
critical temperatures of the confinement/deconfinement phase transition 
and of the chiral-symmetry breaking/restoration phase transition still 
coincide and that the common transition shifts to larger temperatures 
as $\mu_5$ increases. It is shown that this conclusion remains true
in the limit of zero quark mass. This result is in contradiction to the 
results within different effective models of 
QCD \cite{Fukushima:2010fe,Chernodub:2011fr,Gatto:2011wc,Chao:2013qpa,Yu:2014sla} where the critical temperature is said to decrease as $\mu_5$ increases. 

We concede that we study SU(2) QCD with $N_f=4$ quarks which is different 
from what is considered in the effective models, where SU(3) QCD with $N_f=2$
quarks is considered. It would therefore be very interesting for a closer 
comparison to examine the
results of these models by a lattice simulation.

It is likely, however, that the contradictions are rather a consequence of 
the fact that the critical temperature is not a universal parameter but 
crucially depends on the structure and parameters of the  effective models. 
It is unclear to what extent they describe the actual behavior of finite 
temperature QCD. It should be also noted that some predictions obtained in 
different effective models contradict among each others. For instance, the 
dependence of the chiral condensate on $\mu_5$ obtained in 
papers \cite{Fukushima:2010fe,Chernodub:2011fr} is opposite 
to the result obtained in paper \cite{Andrianov:2013qta}. The latter paper 
claims that the chiral condensate grows with the chiral chemical potential, 
which is in agreement with our results. 
 
In addition to the dependence of the critical temperature on the chiral 
chemical potential, some effective models predict that -- beginning from 
some critical value of $\mu_5^c$ ($\mu_5^c \approx 400$ MeV in Ref. 
\cite{Fukushima:2010fe}, $\mu_5^c=50$ MeV in Ref. \cite{Chernodub:2011fr}) --
the transition from the hadronic to the plasma phase 
turns into a first order transition. In our simulations we see that the 
transition becomes sharper as we increase $\mu_5$, but we don't see a 
first order phase transition up to $\mu_5=3345$ MeV, which would manifest 
itself as discontinuity in Polyakov loop and chiral condensate at a 
sufficiently large spatial extension of the lattice. In this connection 
a finite size scaling analysis would be valuable but is outside the scope 
of this paper. 

Besides within effective models, the phase diagram of QCD in the 
$(\mu_5, T)$--plane was studied in papers \cite{Wang:2015tia, Xu:2015vna}
in a framework of Dyson-Schwinger equations. The authors of these papers 
found that the critical temperature rises with $\mu_5$ and the ``phase 
transition'' actually is always a crossover. 
These results are corroborating the results of our paper. 

 We would like also to mention the  paper \cite{Hanada:2011jb}. In this
 paper the authors address the question of universality of phase diagrams in QCD and QCD-like
 theories through the large-$N_c$ equivalence. Using the results of this paper one can show that at
 large-$N_c$ and the chiral limit there is equivalence between QCD phase diagram at finite $\mu_5$ and QCD
 phase diagram at finite isospin chemical potential $\mu_I$. The chiral condensate of QCD with $\mu_5 \neq 0$ 
 can be mapped to the  pion condensate of QCD with $\mu_I \neq 0$. In latter theory the pion condensate and
 critical temperature $T_c$ of pion condensation increase with $\mu_I$. Despite the fact that we considered $N_c=2$ one can
 expect that chiral condensate and $T_c$ in our theory also increase with with $\mu_5$. We believe this is one more
 fact in favour of our results.

\section*{Acknowledgements}

The authors are grateful to V. I. Zakharov and V. G. Bornyakov for 
interesting and stimulating discussions. The authors are grateful
to N. Yamamoto who paid our attention to universality in large-$N_c$ QCD like theories.
The simulations were performed at GPUs of the K100 supercomputer
of the Institute of Applied Mathematics of the Russian Academy of
Sciences in Moscow and at GPUs of the Particle Phenomenology group
at the Institute of Physics of the Humboldt University Berlin.
The work was supported by Far Eastern Federal University, 
by RFBR grants 13-02-01387-a, 14-02-01185-a, 15-02-07596-a, by a grant of the president of the RF, MD-3215.2014.2,
and by a grant of the FAIR-Russia Research Center.

\appendix

\section{Ultraviolet divergences in the chiral condensate}

The fermion propagator including the chiral chemical potential for "naive" 
lattice fermions can be written in the following form 
\beq
S^{\alpha \beta}(x,y) &=& \frac {\delta^{\alpha \beta}} {L_t L_s^3} \sum_{ \{p\} } \sum_s  e^{i p (x-y)} \frac {-i \sum_{\mu} \gamma_{\mu} sin (p_{\mu}) + m + \mu_5 \gamma_0 \gamma_5 }
 {sin^2 (p_0) + (|p|-s \mu_5)^2 +m^2} \times P(s),
\\ \nonumber
P(s) &=& \frac 1 2 \biggl ( 1 -  i s \sum_i \frac {\gamma_{i} sin (p_{i}) } {|p|} \gamma_0 \gamma_5  \biggr ),~~~i=1,2,3,
\\ \nonumber
|p|^2 &=& sin^2 (p_1) + sin^2 (p_2) + sin^2 (p_3),
\\ \nonumber
p_i &=& \frac {2 \pi} {L_s} n_i,~~ i=1,2,3,~~ n_i=0,...,L_s-1, \\ \nonumber
p_4 &=& \frac {2 \pi} {L_t} n_4 + \frac {\pi} {L_t},~~~ n_4=0,...,L_t-1.
\eeq
Here $m$ and $\mu_5$ are mass and chiral chemical potential in lattice units, 
$\alpha, \beta$ are color indices, the sum is taken over all possible values 
of $(n_1, n_2, n_3, n_4)$ and $s= \pm 1$. 

In the limit $L_s, L_t \to \infty$ the condensate per one fermion flavour can be written as 
\beq
\langle \bar \psi \psi \rangle =  \frac 1 {16} Tr [S(x,x)]=   
\frac  m 4  \int_{-\pi}^{\pi} \frac {d^4 p} {(2 \pi)^4} 
\sum_{s} \frac 1 {sin^2 (p_0) + (|p|-s \mu_5)^2 +m^2} .
\label{chiral_condensate}
\eeq
To calculate the integral in formula (\ref{chiral_condensate}) we use the 
standard approach, which separates the main divergence from the rest of the 
integral
\beq
\label{integral_chiral_condensate}
\int_{-\pi}^{\pi} \frac {d^4 p} {(2 \pi)^4} \sum_{s} \frac 1 {sin^2 (p_0) + (|p|-s \mu_5)^2 +m^2} = \int_{-\pi}^{\pi} \frac {d^4 p} {(2 \pi)^4} \frac 2 {sin^2 (p_0) + |p|^2 +m^2} + \\ \nonumber
\int_{-\pi}^{\pi} \frac {d^4 p} {(2 \pi)^4} \sum_{s} \biggl ( \frac 1 {sin^2 (p_0) + (|p|-s \mu_5)^2 +m^2} - 
 \frac 1 {sin^2 (p_0) + |p|^2 +m^2} \biggr ). 
\eeq
The first term in this expression is just the loop 
integral for $\langle \bar \psi \psi \rangle$ without chiral chemical 
potential. 
The expression for this integral in the continuum limit $a \to 0$ can be 
found in \cite{Capitani:2002mp}
\beq
\int_{-\pi}^{\pi} \frac {d^4 p} {(2 \pi)^4} \frac 1 {sin^2 (p_0) + |p|^2 +m^2}  = 0.619734 + m^2 \biggr (  \frac {log (m^2)} {\pi^2} -0.345071 \biggl ) 
+ O(m^4) .
\label{chiral_condensate_m}
\eeq 
The  second term in (\ref{integral_chiral_condensate}) is proportional to  
$\mu_5^2$ and contains a logarithmic divergence in the continuum limit. 
This divergence can be calculated using saddle-point method. 
To calculate the second term in (\ref{integral_chiral_condensate}) we 
subtract the leading divergence and calculate the rest of the integral 
in the limit $a \to 0$. 
The result of the calculation can be written as follows
\beq
\int_{-\pi}^{\pi} \frac {d^4 p} {(2 \pi)^4} \sum_{s} \biggl ( \frac 1 {sin^2 (p_0) + (|p|-s \mu_5)^2 +m^2} - 
 \frac 1 {sin^2 (p_0) + |p|^2 +m^2} \biggr ) = \nonumber \\ = 
\mu_5^2 \biggl ( -4  \frac {log (m^2)} {\pi^2} + 0.671036 \biggr ) 
+ O(m^4, \mu_5^2 m^2, \mu_5^4)  .
\label{chiral_condensate_mu5}
\eeq
We have checked the last formula numerically. Introducing a physical mass 
$m_p a = m$ and a physical chiral chemical potential $\mu_5^p a= \mu_5$
one can write the expression for the condensate in physical units as
\beq
\langle \bar \psi \psi \rangle_p = 0.309867 \frac {m_p} {a^2} + m_p^3 \biggl ( \frac {log(m_p a)} {\pi^2} - 0.172536 \biggr ) + 
m_p (\mu_5^p)^2 \biggl (-2 \frac {log(m_p a)} {\pi^2} + 0.167759  \biggr ) .
\label{condensate_ph}
\eeq
From the last formula one sees that at one-loop level the inclusion of a
non-zero chiral chemical potential leads to an additional logarithmic 
divergence. 
We believe that this conclusion persists if one takes into account radiative 
corrections to formula (\ref{condensate_ph}). 
To see this, consider some Feynman graph of radiative corrections to 
formula (\ref{condensate_ph}). 
The expansion of this graph in powers of $\mu_5$ can be reduced to the 
inclusion of a dimension-3 operator to the graph that diminishes the power 
of divergence by one unit per one power of $\mu_5$. Expansion of the chiral 
condensate in powers of $\mu_5$ contains only even powers. The main 
divergence of the chiral condensate is $\sim 1/a^2$.
So, the next to leading term in $\mu_5$ expansion is two powers smaller 
than $\sim 1/a^2$, i.e. it is at most logarithmically divergent.

\begin{figure}[h]
 \includegraphics[width =5.5cm, angle =270]{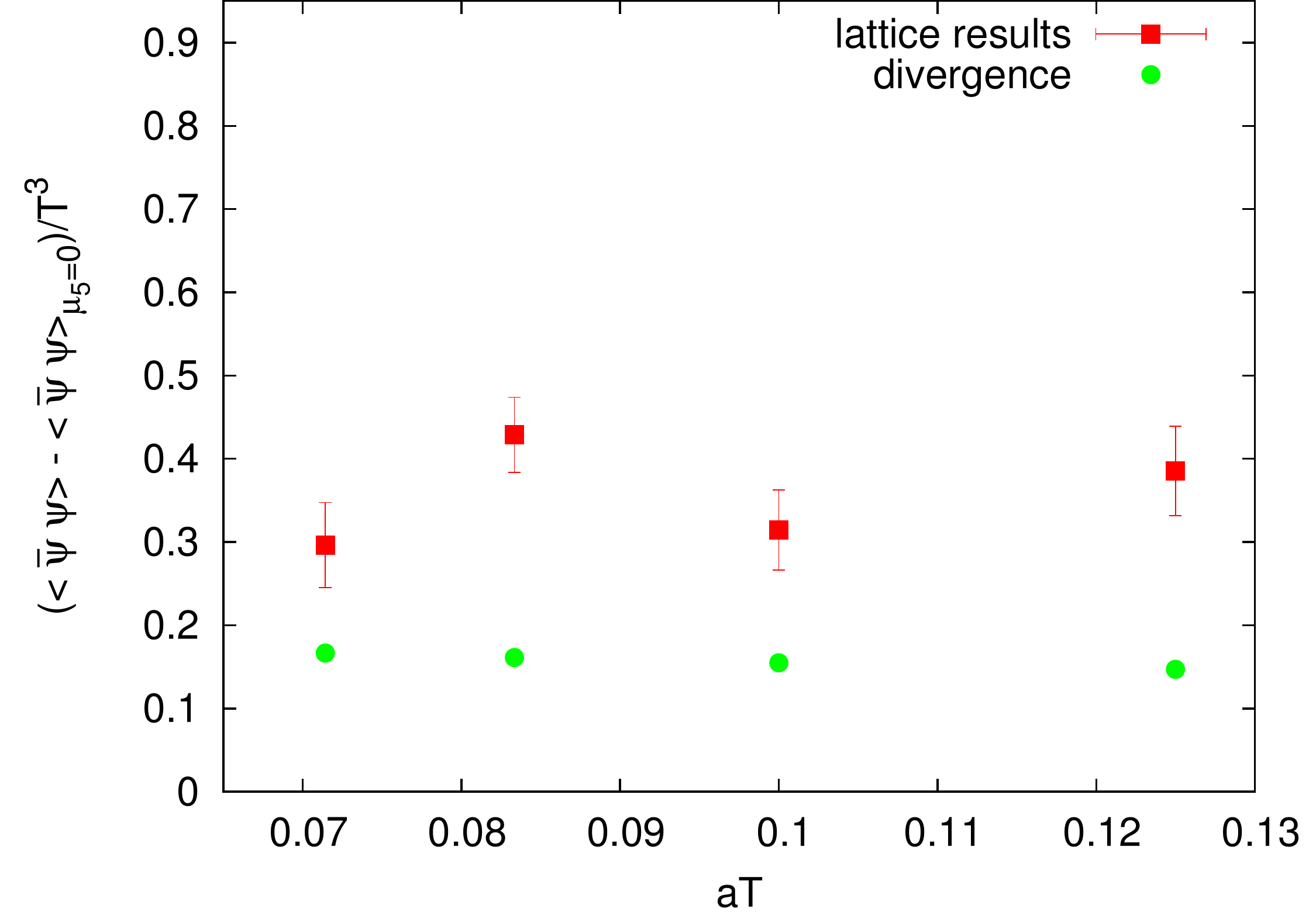}\\
 \caption{ The difference between chiral condensates at $\mu_5=0$  MeV and  $\mu_5=330$  MeV for different lattice spacings ($T=270$ MeV). 
Square points correspond to lattice measurements, circle points correspond to  the contribution of logarithmic 
divergence (\ref{condensate_ph}) to the difference.
}
 \label{condensate_div}
\end{figure}

Further it is important to understand how the additional divergence connected 
to non-zero $\mu_5$ effects the results of this paper. To estimate this 
we fixed the physical values of temperature $T=270$ MeV, quark mass $m_p=33$ 
MeV and measured the condensate $\langle \bar \psi \psi \rangle$ in the 
deconfinement phase for $\mu_5=0$ and $\mu_5=330$ MeV for different values 
of lattice spacing $a$. In particular, we took the following 
lattice parameters: $16^3 \times 8~\beta=1.9500$, 
$20^3 \times 10~\beta=2.0047$, 
$24^3 \times 12~\beta= 2.0493$ and $28^3 \times 14~\beta= 2.0870$.
The difference between the measured chiral condensates at zero and non-zero 
$\mu_5$ for different lattice spacings is shown in Fig.\ref{condensate_div}. 
In addition to the lattice measurement, in Fig.\ref{condensate_div} we plot 
the contribution of logarithmic divergence (\ref{condensate_ph}) to the 
difference 
$\langle \bar \psi \psi \rangle - \langle \bar \psi \psi \rangle_{\mu_5=0}$.

From Fig.\ref{condensate_div} one sees that uncertainty of the calculation 
doesn't allow to confirm the logarithmic type of the divergence. However, 
the variation of the lattice results with variation of the lattice spacing 
is very slow what allows us to assume that there are no other divergences 
different from the logarithmic one. Note also that the value of the 
$\mu_5 \neq 0$ contribution to the chiral condensate obtained from the 
one-loop expression for the logarithmic divergence is by a factor 2-3 
smaller than the lattice results. The difference can be attributed to a
non-zero temperature effect and radiative corrections. This fact allows 
one to state that the renormalization effect of the one-loop expression 
for the logarithmic divergence (\ref{condensate_ph}) is not very large. 
So one can use it to estimate the contribution of ultraviolet logarithmic 
divergence in the confinement phase.

The calculation shows that the characteristic values of the difference 
$\langle \bar \psi \psi \rangle - \langle \bar \psi \psi \rangle_{\mu_5=0}$  
in the confinement phase for different values of lattice parameters used 
in this paper are approximately by one order of magnitude larger than 
the additional ultraviolet logarithmic divergence due to $\mu_5 \neq 0$ 
estimated according to formula (\ref{condensate_ph}). 
For instance, from Fig. \ref{fig:small} one can see that at temperature 
$T=200$ MeV (close to the phase transition at $\mu_5=0$)  
the difference 
$(\langle \bar \psi \psi \rangle_{\mu_5=300 \mbox{MeV}} - \langle \bar \psi \psi \rangle_{\mu_5=0})/T^3 \approx 2$.
At the same time the additional ultraviolet divergence according to formula 
(\ref{condensate_ph}) gives 
$(\langle \bar \psi \psi \rangle_{add})/T^3 \approx 0.1$. 
Note also that the position of the phase transition 
manifests itself as a peak of the susceptibility. 
Evidently, there is no peak due to the additional ultraviolet 
divergence. All this allows us to state that the conclusions obtained in 
this paper are not affected by the additional ultraviolet $\mu_5 \neq 0$ 
divergence in the chiral condensate.

\section{Ultraviolet divergences in the Polyakov loop}

At the leading order approximation in the strong coupling constant $g$ 
the Polyakov loop can be written in the following form
\beq
L = 1 - \beta \frac {g^2} {4} \int_0^{\beta} d \tau D^{aa}_{00}(\tau, \vec 0) 
=1 - \beta \frac {g^2} 4 \int \frac {d^3 q} {(2 \pi)^3}  {\tilde D}^{aa}_{00}(0, \vec q),
\label{Polyakov}
\eeq
where $D^{ab}_{00}(t, \vec x)$ and ${\tilde D}^{aa}_{00}(q_0, \vec q)$ 
represent the propagator of the temporal components of the gluon field
in coordinate and momentum space, correspondingly, 
$\beta$ is inverse temperature, and the summation over color index $a$ is 
assumed. From eq. (\ref{Polyakov}) it is clear that additional divergences 
connected to non-zero $\mu_5$ at one-loop level can appear from the fermion 
self-energy part of the gluon propagator. The study of the divergences 
with "naive" fermions in the self-energy part of the propagator is cumbersome. 
For this reason we use momentum cut of regularization procedure 
to study the divergences in the Polyakov loop. 

\begin{figure}[h]
 \includegraphics[width =5.5cm, angle =270]{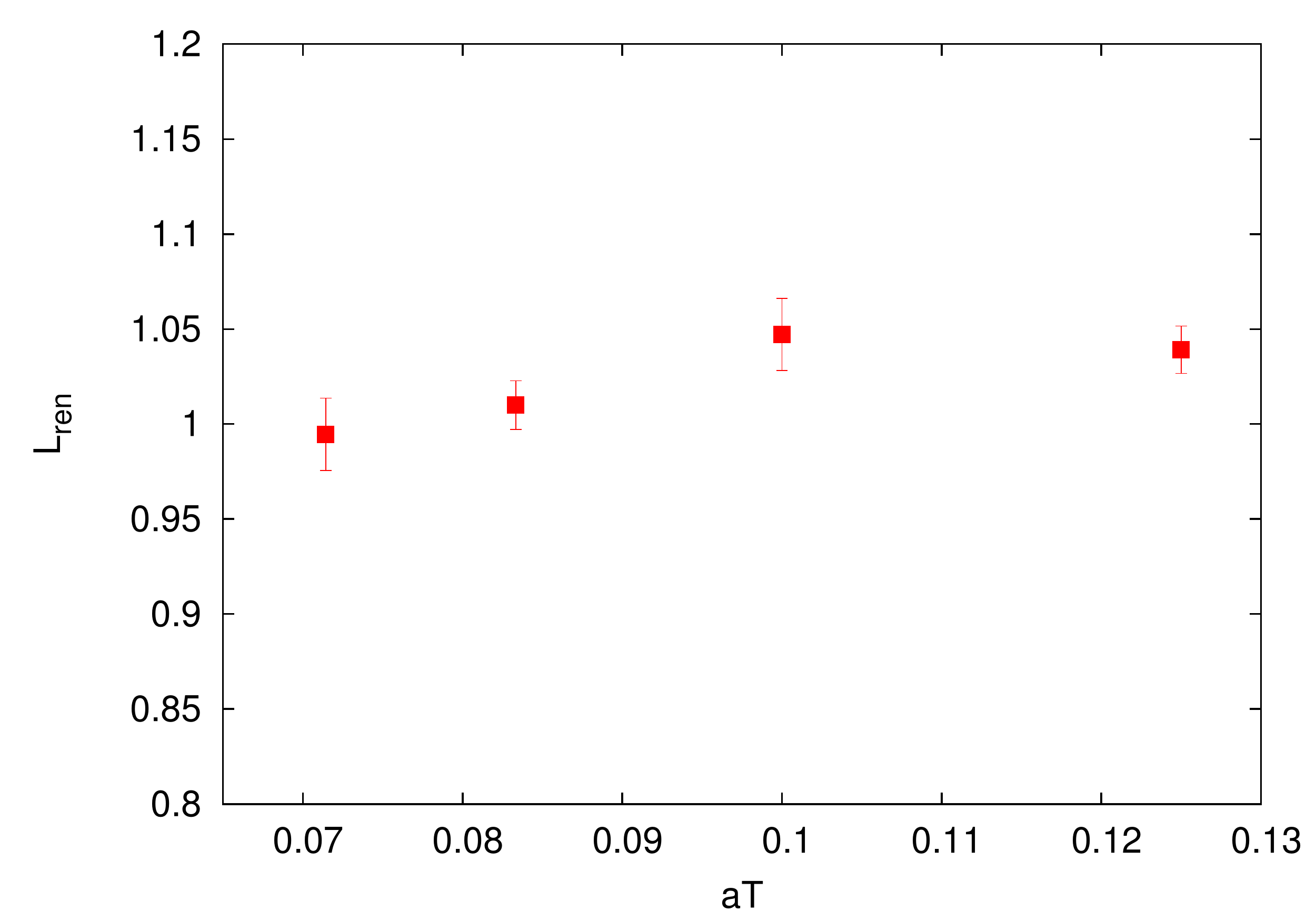}\\
 \caption{ Renormalized Polyakov loop (\ref{polyakov_ren}) as a function of lattice spacing ($T=270$ MeV). }
 \label{polyakov}
\end{figure}

The fermion contribution to the one-loop self-energy of the gluon propagator 
${\tilde D}^{ab}_{00}$ can be written in the following form
\beq
\Pi_{00} (\vec q) = \frac 1 4 \sum_{s_1, s_2} \int^{\Lambda} \frac {d^4 k} {(2 \pi)^4} \frac {-k_0^2 + (\mu_5 - s_1 |\vec k|) (\mu_5 - s_2 |\vec k+\vec q|) +m^2}
{(k_0^2 + (|\vec k|-s_1 \mu_5)^2 +m^2) (k_0^2 + (|\vec k + \vec q |-s_2 \mu_5)^2 +m^2)} \biggl ( 1 + s_1 s_2 \frac {\vec k \cdot (\vec k + \vec q)} {|\vec k| |\vec k + \vec q|} 
 \biggr )
\label{self_energy}
\eeq
where $\vec q$ is the external momentum, $s_1,s_2 = \pm 1$. 

To proceed we expand equation (\ref{self_energy}) in $\mu_5^2$. Evidently 
each new term in this expansion 
diminishes the power of divergence by two units. That means that beginning 
from $\mu_5^4$ term the series is ultraviolet 
convergent. The first term $(\Pi_{00}(\vec q)|_{\mu_5= 0})$ is just the 
expression for the self-energy without chiral chemical potential. 
So, to calculate the divergence which results from nonzero $\mu_5$ one needs 
to study only the divergence of the second term. Taking 
the derivative of equation (\ref{self_energy}) with respect to 
$\mu_5^2$ we get
\beq
\label{B3}
\Pi_{00} (\vec q) - \Pi_{00} (\vec q)|_{\mu_5=0} = 4 \mu_5^2 \int \frac {d^4 k} {(2 \pi)^4} \biggl (
\frac {1} {(k_0^2 +|\vec k|^2 +m^2) (k_0^2 +|\vec k + \vec q|^2 +m^2)} -  \\ \nonumber
-\frac {2 m^2 +4 |k+q|^2 + 6 \vec k \cdot (\vec k + \vec q) -2 k_0^2} {(k_0^2 +|\vec k|^2 +m^2)^2 (k_0^2 +|\vec k + \vec q|^2 +m^2)} +
\frac { 8 |\vec k|^2 ( m^2 + \vec k \cdot (\vec k + \vec q) - k_0^2)} {(k_0^2 +|\vec k|^2 +m^2)^3 (k_0^2 +|\vec k + \vec q|^2 +m^2)} + \\ \nonumber
+\frac {4 (m^2-k_0^2) (\vec k \cdot (\vec k + \vec q)) + 
4 |\vec k|^2 |\vec k +\vec q|^2} {(k_0^2 +|\vec k|^2 +m^2)^2 (k_0^2 +|\vec k + \vec q|^2 +m^2)^2} 
\biggr ) +O(\mu_5^4)
\eeq
One can easily check that this expression is free from ultraviolet divergence. 
Moreover, a calculation shows that at large $\vec q^2$ expression (\ref{B3}) 
behaves as  
$\Pi_{00} (\vec q) - \Pi_{00} (\vec q)|_{\mu_5=0} \sim \mu_5^2 (c_1 \log {\vec q^2} +c_2)$, 
where $c_1, c_2$ are some constants. So, at one-loop level there is no 
ultraviolet divergence in the Polyakov loop which results from non-zero 
chiral chemical potential. Similarly to the previous section, one can argue 
that there is no ultraviolet divergence in the $\mu_5 \neq 0$ contribution 
to the Polyakov loop at higher loops. 

In order to numerically study the role of divergences in the Polyakov loop 
due to non-zero chiral chemical potential,
we used the following definition of renormalized Polyakov loop
\beq
L_{ren} = \frac {L} {L_{\mu_5=0}}, 
\label{polyakov_ren}
\eeq
where $L$ is the Polyakov loop and $L_{\mu_5=0}$ is the Polyakov loop at 
the same lattice but with $\mu_5=0$.
Evidently $L_{ren}$ doesn't contain divergences 
that are usual for simulations with zero chiral chemical potential. 
Similarly to the previous section we fixed the physical values of 
temperature $T=270$ MeV, quark mass $m_p=33$ MeV and measured 
the $L_{ren}$ in the deconfinement phase for $\mu_5=330$ MeV and 
different values of lattice spacing $a$.
The result of this measurement is shown in Fig. \ref{polyakov}

From this figure one sees that the renormalized Polyakov loop
$L_{ren}$ is consistent with a constant in the region of
lattice spacings investigated. In the same region the unrenormalized
Polyakov loop $L$ changes by a factor of two. From these facts we conclude
that there is no additional ultraviolet divergence in  Polyakov loop due 
to non-zero chiral chemical potential.

\end{document}